\documentclass[%
	aps,
	prb,
	amsmath,
	amssymb,
	reprint,
	superscriptaddress,
	showpacs,
	floatfix,
]{revtex4-1}

\usepackage{graphicx}% Include figure files
\usepackage{dcolumn}% Align table columns on decimal point
\usepackage{bm}% bold math

\begin{document}

\title{Reconstruction of Tip-Surface Interactions with Multimodal Intermodulation Atomic Force Microscopy}

\author{Stanislav S. Borysov}
\email[]{borysov@kth.se}
%\thanks{}
\affiliation{Nanostructure Physics, KTH Royal Institute of Technology, Roslagstullsbacken 21, SE-106 91 Stockholm, Sweden}
\affiliation{Nordita, KTH Royal Institute of Technology and Stockholm University, Roslagstullsbacken 23, SE-106 91 Stockholm, Sweden}

\author{Daniel Platz}
\affiliation{Nanostructure Physics, KTH Royal Institute of Technology, Roslagstullsbacken 21, SE-106 91 Stockholm, Sweden}

\author{Astrid S. de Wijn}
\affiliation{Department of Physics, Stockholm University, SE-106 91 Stockholm, Sweden}

\author{Daniel Forchheimer}
\affiliation{Nanostructure Physics, KTH Royal Institute of Technology, Roslagstullsbacken 21, SE-106 91 Stockholm, Sweden}

\author{Eric A. Tol\'en}
\affiliation{Intermodulation Products AB, Vasav\"agen 29, Solna SE-169 58, Sweden}

\author{Alexander V. Balatsky}
\affiliation{Nordita, KTH Royal Institute of Technology and Stockholm University, Roslagstullsbacken 23, SE-106 91 Stockholm, Sweden}
\affiliation{Theoretical Division and Center for Integrated Nanotechnologies, Los Alamos National Laboratory, Los Alamos, NM 87545, USA}

\author{David B. Haviland}
\affiliation{Nanostructure Physics, KTH Royal Institute of Technology, Roslagstullsbacken 21, SE-106 91 Stockholm, Sweden}
\email[]{haviland@kth.se}

\date{\today}

\begin{abstract}
We propose a theoretical framework for reconstructing tip-surface interactions using the intermodulation technique when more than one eigenmode is required to describe the cantilever motion.  Two particular cases of bimodal motion are studied numerically: one bending and one torsional mode, and two bending modes. We demonstrate the possibility of accurate reconstruction of a two-dimensional conservative force field for the former case, while dissipative forces are studied for the latter.
\end{abstract}

%\keywords{}

\maketitle

\section{Introduction} 
Atomic force microscopy \cite{AFM1} (AFM) has become one of the most important tools for the study of nanometer-scale surface properties of a wide range of materials. The initial goal of AFM was surface topography imaging which was performed by scanning a cantilever with a sharp tip over a surface while keeping its deflection constant. It was later realized that the reconstruction of tip-surface interactions was possible and that these interactions contain valuable information about material properties \cite{forceIterpret1,forceIterpret2}. One of the first reconstruction methods was based on measurement of the quasi-static bending of the cantilever beam as its base was slowly moved toward and away from the surface. Two drawbacks of this method are the slow speed of measurement and the lack of ability to reconstruct dissipative forces which are always present in tip-surface interactions due to non-elastic deformations of the sample, breaking chemical bonds, or other irreversible processes \cite{nanoDiss1,nanoDiss2,nanoDiss3,nanoDiss4,nanoDiss5}.

The development of dynamic AFM opened new pathways for a more profound study of tip-surface interactions. In dynamic AFM the cantilever is treated as an underdamped oscillator (high $Q$-factor) driven at a resonance where the response to external forces is enhanced by a factor $Q$. A large number of methods for determining the tip-surface interaction have been devised \cite{dynAFM1,dynAFM2,dynAFM3,dynAFM4,dynAFM5,dynAFM6,dynAFM7},
some making use of amplitude or frequency modulation
 \cite{like_imafm0,like_imafm1,like_imafm2,like_imafm3,like_imafm4,like_imafm5,like_imafm6,like_imafm7,like_imafm8,like_imafm9}.
In this paper, we consider intermodulation AFM (ImAFM) \cite{imafm1} which is unique in its ability to rapidly extract a large amount of information about tip-surface interactions in a simple and convenient way \cite{imafm2,imafm3}. With ImAFM we have the possibility of high resolution surface property mapping at interactive scan speeds and reconstruction of the tip-surface interaction at each pixel.

\begin{figure}
\centering
\includegraphics[width=75mm]{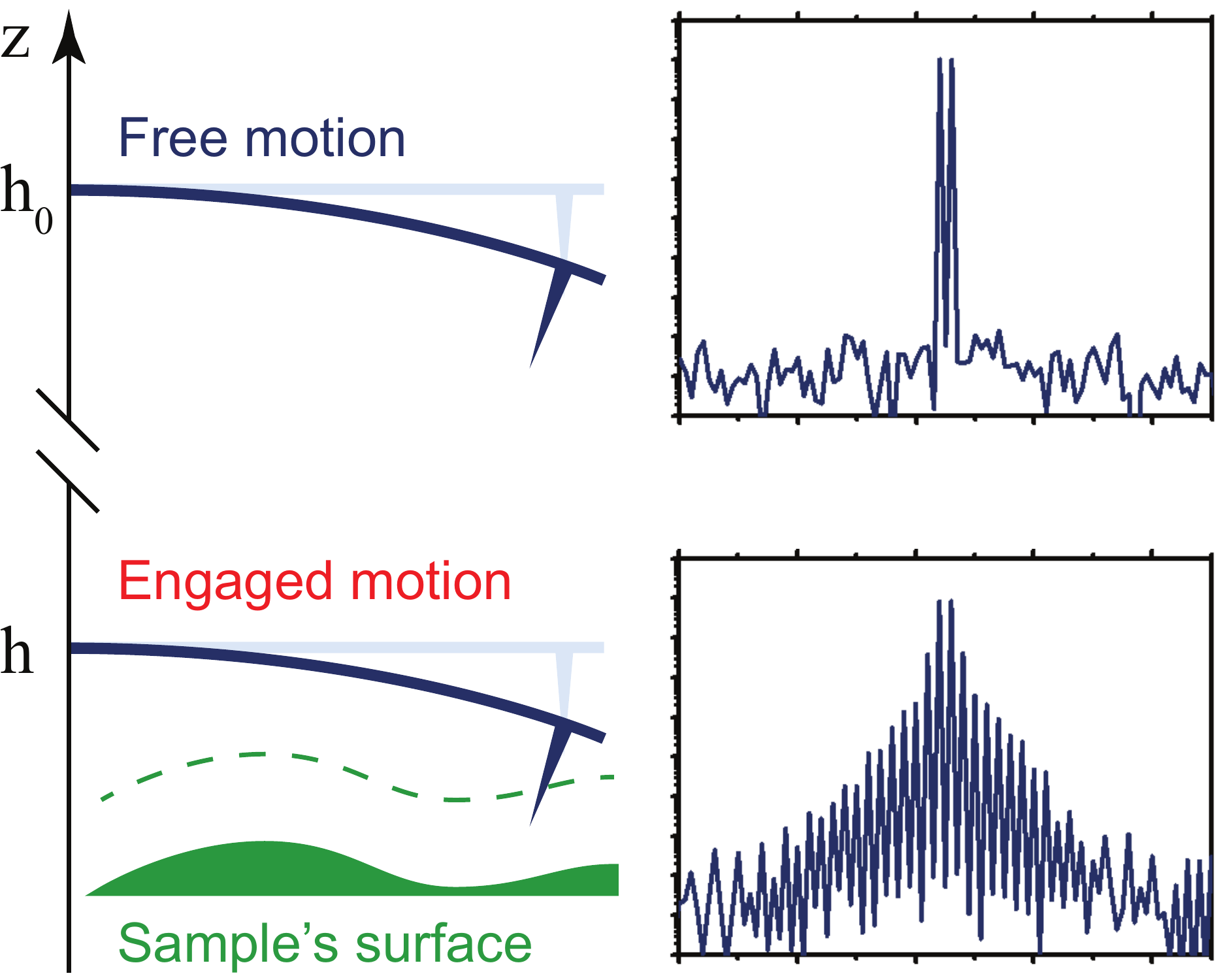}
 \caption{\label{fig:imafm}(Color online) Appearance of new frequencies in the spectrum of the tip motion in ImAFM due to the nonlinearities in tip-surface interaction. The dashed line shows the characteristic range of the tip-surface force.}
\end{figure}

The main idea underlying ImAFM is to use the nonlinear tip-surface forces to create high-order intermodulation of discrete tones in a frequency comb \cite{combtones}. The method can be generally applied to any resonator subject to a nonlinear force when it is driven with at least two frequencies $\omega_{d1}$ and $\omega_{d2}$.  The nonlinear response  in frequency domain occurs not only at the drive frequencies and their harmonics $n\omega_{d1}$ and $m\omega_{d2}$, where $n$ and $m$ are integers, but also at their linear combinations $\omega=n\omega_{d1}+m\omega_{d2}$ or intermodulation products. Due to the signal enhancement near resonance and finite detector noise of the measurement, a typical spectrum of cantilever motion can be obtained only in narrow frequency band near resonance. Concentrating as many intermodulation products as possible in this resonant detection band results in more information for reconstruction of the nonlinear forces (fig.~\ref{fig:imafm}).

The sensitivity enhancement of dynamic AFM occurs not only at one single eigenmode, but also at several eigenmodes simultaneously. Measurement of the sample response at higher frequencies allows for better material imaging contrasts and improves accuracy of the tip-surface force reconstruction \cite{multi1,multi2,multi3,multi4,multi5,multi6,multi8}. If we also excite a torsional resonance of the cantilever we open up the possibility of measuring lateral forces acting on the tip \cite{tors0,tors1,tors2}. Additionally, when driving the cantilever at more than one eigenmode, new frequency bands become available for collecting intermodulation products, resulting from the nonlinear force which couples multiple eigenmodes \cite{multi10}. While ImAFM has been well studied for the case of one flexural mode \cite{imafm2,imafm3,imafm4,imafm5,imafm6}, the multi-eigenmode problem remains open. In the current investigation we explore the possibility of reconstructing two-dimensional tip-surface force fields using two collinear eigenmodes (e.g. two flexural or two torsional modes) as well as two orthogonal modes (e.g. one flexural and one torsional mode).

The paper is organized as follows: in section \ref{sec:model_assumptions} we consider general properties of the cantilever model and tip-surface interactions. In section \ref{sec:imafm} we recount the main principles of ImAFM, demonstrate how to obtain intermodulation spectra of tip-surface forces, and develop an extension of the spectral fitting method \cite{imafm2,imafm3} of force reconstruction for the multimodal case. In Section \ref{sec:results} numerical results for the reconstruction of two-dimensional conservative and dissipative forces are presented. Section \ref{sec:conclusions} concludes with a discussion and summary.

%-------------------------------------------------------------------------------
\section{\label{sec:model_assumptions}Model}

\begin{figure*}
\centering
\includegraphics[width=150mm]{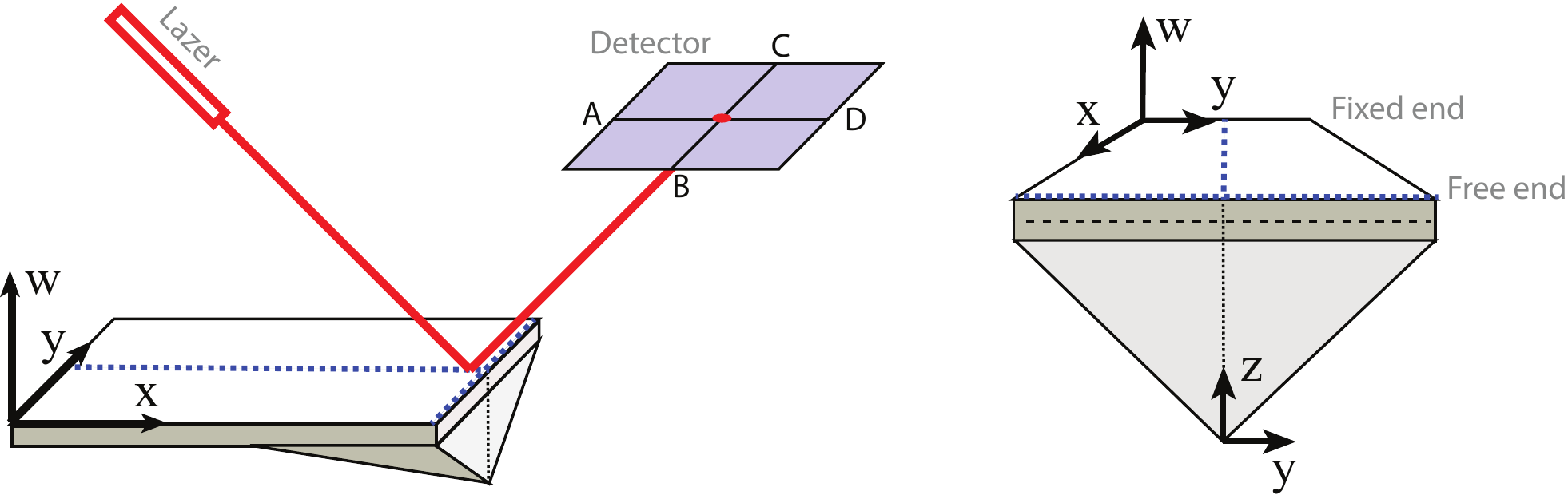}
 \caption{\label{fig:cantilever}(Color online) Schematic representation of the two-dimensional cantilever and detection principle in AFM.}
\end{figure*}

In order to proceed with the multimodal problem we should start from some general discussion of cantilever dynamics \cite{gen1_cantDyn,gen2_cantVib}. If we are interested in studying only flexural modes, one-dimensional Euler-Bernoulli beam theory \cite{bern1,bern2,bern3,bern4} is sufficient.  Incorporating torsional modes requires the cantilever to be regarded as a two-dimensional object.  Many theories have been developed describing the continuum mechanics of two-dimensional plates \cite{plate1_love,plate2_timosh,plate3_review,plate4_andtors}. A general description of an arbitrary two-dimensional cantilever (fig.~\ref{fig:cantilever}) is given by the governing equation 
\begin{equation}
	\left(\mathcal{G}_{xy}+\mathcal{G}_t\right)\left[w\left(x,y,t\right)\right] = F + \mathrm{f}
	\label{supereqModel}
\end{equation} 
with an appropriate set of boundary conditions. 
Here $x$ and $y$ are the two space coordinates, $t$ is time, $w(x,y,t)$ is a deflection normal to the $x - y$ plane of the plate at rest, $\mathcal{G}_{xy}$ is a space coupling operator which corresponds to the two-dimensional mathematical model used to determine the stresses and deformations in thin plates, and $\mathcal{G}_t$ is a time evolution operator which represents inertia and damping. $F$ and $\mathrm{f}$ are a scalar quantities which are projections on to $w$, of the two-dimensional vector force field acting on the tip, and the drive, respectively. In general, the force $F$ can depend explicitly on the deflection $w$, its velocity $\dot w$, past trajectories $\left.\{w,\dot w\}\right|_{-\infty}^{t}$ and time $t$. In this paper we restrict ourselves to $F$ which are not dependent on past trajectories. We consider only the case of small deflections, so $\mathcal{G}_{xy}$ and $\mathcal{G}_t$ are linear. For example, within the Kirchhoff-Love plate theory \cite{plate1_love} the space operator for a homogeneous cantilever reads
\begin{equation}
	\mathcal{G}_{xy} := \frac{2h^3E}{3(1-\nu^2)}\left(\frac{\partial^4}{\partial x^4}+2\frac{\partial^4}{\partial x^2\partial y^2}+\frac{\partial^4}{\partial y^4}\right)
	\label{kirchhoff}
\end{equation}
where $2h$ is the cantilever thickness, $E$ the Young's modulus and $\nu$ the Poisson's ratio. The theory assumes that a mid-surface plane can be used to represent a three-dimensional plate in a two-dimensional form. The time evolution operator $\mathcal{G}_t$ consists of an inertial term, and for the case of a homogeneous viscous media, a linear damping term 
\begin{equation}
	\mathcal{G}_t:=2\rho h\frac{\partial^2}{\partial t^2}+2\gamma\frac{\partial}{\partial t}
	\label{timeoperLove}
\end{equation}
where $\rho$ is the cantilever density and $\gamma$ a damping coefficient. 

The linearity of $\mathcal{G}_{xy}$ and $\mathcal{G}_t$ gives a dynamics of the two-dimensional cantilever which is well approximated by a system of differential equations for the generalized coordinates $q_i$ representing deflections of its normal modes (see appendix \ref{app:coord} for a derivation)
\begin{equation}
	k_i\left(\frac{1}{\omega_i^2}\ddot q_i + \frac{1}{Q_i\omega_i}\dot q_i + q_i\right) = F_{i}\left(t\right) + \mathrm{f}_i\left(t\right)
	\label{worksys}
\end{equation}
Here each generalized coordinate $q_{i}$ has a stiffness $k_{i}$, resonant frequency $\omega_{i}$ and quality factor $Q_{i}$. $F_{i}(t)$ is the projection of the nonlinear force and $\mathrm{f}_{i}(t)$ is the projection of the drive force on to the $i^{\mathrm{th}}$ eigenmode. The problem of mapping the eigenmode coordinates $q_{i}$ onto the physical position of the tip $\mathbf{r}\equiv(z,y)^\intercal$, (hereafter $\intercal$ denotes transpose) as well as the relationship between $F_i$ and the actual vector force field acting on the tip $\mathbf{F}_{ts}(\mathbf{r},\dot{\mathbf{r}})$, is determined by the cantilever and tip geometry as briefly discussed in section \ref{sec:results}. 

%-------------------------------------------------------------------------------
\section{\label{sec:imafm}Reconstructing force from intermodulation with multiple eigenmodes}

We recall the basic steps to get information about $F_{i}$ in the frequency domain using the intermodulation technique. Firstly, we excite one or several generalized coordinates (\ref{worksys}) with the drive $\mathrm{f}_{i}$ and measure its motion spectrum at the height $h_0$, far from the sample surface, so $\left.\hat{F}_{i}\right|_{h_0}\equiv 0$ for all $t$. This free oscillation spectrum is denoted
\begin{equation}
	\left.\hat{q}_{i}\right|_{h_0}=\chi_{i}\hat{\mathrm{f}}_{i}
	\label{sys_imh0}
\end{equation}
where the hat represents the discrete Fourier transform (DFT)
\begin{equation}
\hat q(\omega) \equiv \mathcal{F}[q(t)] := \sum\limits_{m=-\infty}^{\infty}q(t)e^{-2\pi\imath m\,\delta\omega t}  %\int\limits_{-\infty}^{+\infty}f(t)e^{-2\pi i\omega t}\,\mathrm{d}t
	\label{ft}
\end{equation}
and the linear response function $\chi_{i}$ is introduced 
\begin{equation}
\chi_{i}(\omega) = k_{i}^{-1}\left[1 + \frac{\imath}{Q_{i}}\left(\frac{\omega}{\omega_{i}}\right) -\left(\frac{\omega}{\omega_{i}}\right)^2\right]^{-1}
	\label{chi}
\end{equation}
Using pairs of drive tones with the same amplitude $A_{i}$ which are separated by $\delta\omega$ and placed close to each resonance $n_i\delta\omega<\omega_i<(n_i+1)\delta\omega$, where $n_i$ is an integer,
\begin{equation}
	\mathrm{f}_{i}(t) = A_{i}\left[\cos\left(n_i\delta\omega t\right)+\cos\left((n_i+1)\delta\omega t\right)\right]
	\label{driving_force}
\end{equation}
we obtain a free, linear response spectrum (eq. (\ref{sys_imh0})) consisting components only at the drive frequencies $\omega_{i}\pm\delta\omega/2$ (fig.~\ref{fig:imafm}). Data acquisition in ImAFM should be performed over at least one total period of the drive $T=2\pi/\delta\omega$ while measuring subsequent periods only improves the SNR of the measured spectra \cite{imafm_pr1,imafm_pr2}.

We then we move the cantilever closer to the sample to the engaged height $h$, so the oscillating tip starts to feel interaction with a surface (but not too close, so that there is zero static deflection of the cantilever) and measure the motion spectrum again
\begin{equation}
	\left.\hat{q}_{i}\right|_{h}=\chi_i\left(\hat{F}_{i} + \hat{ \mathrm{f}}_{i}\right)
	\label{sys_imh}
\end{equation}
Finally, the difference between this engaged motion spectrum (\ref{sys_imh}) and free motion spectrum (\ref{sys_imh0}) yields the desired interaction force spectrum
\begin{equation}
	\hat{F}_{i} = \chi_{i}^{-1}\left(\left.\hat{q}_{i}\right|_{h} - \left.\hat{q}_{i}\right|_{h_0}\right) \equiv \chi_{i}^{-1}\Delta\hat{q}_{i}
	\label{hat_f}
\end{equation}
where finite difference operator $\Delta$ is used for short. Thus, we have obtained information about $F_{i}$ in the frequency space representation of the motion and we are in a position to discuss the reconstruction of its dependence on generalized coordinates $q_{i}$ and velocities $\dot q_{i}$. 

Under ideal conditions, given the full difference spectrum $\Delta\hat q_i$ and corresponding response function $\chi_i$, it is possible to find $F_{i}(t)$ as the inverse Fourier transform of $\hat F_{i}(\omega)$ and then trivially recover its coordinate dependence $F_{i}(\{q_{i},\dot q_{i}\})$ using the measured motion $q_i(t)$. However, in real experiments this na\"ive approach fails due to the strong frequency dependence of $\chi_i$ and the measurement limitations imposed by detection noise. In practice, almost all of the spectrum $\hat q_{i}$ is buried under detector noise except for a narrow band near its resonant frequency where the signal-to-noise ratio (SNR) meets the thermal limit. Usually the number of resolvable spectral components $B_{i}$ in a band surrounding each eigenmode resonance is limited to a few dozen, depending on the difference frequency $\delta\omega$ and the forces experienced during the interaction. The use of several ($N$) eigenmodes allows for a lager total number of frequency components $B=\sum_{i=1}^N B_i$ for force reconstruction.

Different methods have been elaborated for the single mode reconstruction problem using this limited amount of response in the resonant detection band \cite{imafm2,imafm3,imafm4,imafm5,imafm6}. Here we develop an extension of the spectral fitting method \cite{imafm2,imafm3,imafm6} for the multimodal problem as it allows for a straightforward generalization without involving any sophisticated concepts \cite{imafm4}. Following the method's main idea, one assumes a tip-surface force in the form of some known model function $\tilde F_{i}(q_1,\dots,q_N;\mathbf{g})$ with a vector $\mathbf{g}=(g_1,\dots\,g_P)^\intercal$ of $P$ unknown parameters. Fitting the calculated spectrum $\hat{\tilde{F}}_{i}(\omega)$ to the measured $\hat F_{i}(\omega)$ (eq. (\ref{hat_f})), we minimize the error function in the frequency domain, in a least square sense
\begin{equation}
\begin{array}{lcl}
	\min\limits_{\mathbf{g}}\hat e_{i} = \hat F_{i} - \hat{\tilde{F}}_{i}\left(q_1,\dots,q_N;\mathbf{g}\right)
\end{array}
\label{errorf}
\end{equation}

The model $\tilde F_{i}$ can be a particular phenomenological expression, for example the van der Waals-Derjagin-Muller-Toporov (vdW-DMT) force \cite{vdW-DMT} or its modifications \cite{vdW-DMT_mod1,vdW-DMT_mod2,vdW-DMT_mod3}. However, in the general case we do not know the exact form of the interaction and should choose some generic function structure, for instance a truncated Taylor expansion in the following polynomial form  
\begin{equation}
	\tilde F_{i}(q_1,\dots,q_N) =
\sum\limits_{i_1=0}^{P_1}\dots \sum\limits_{i_N=0}^{P_N} 
g_{i_1\dots i_{N}}q_1^{i_1}\dots q_N^{i_N} = \mathbf{q}^\intercal\mathbf{g}
\label{poly_general}
\end{equation}
Here $P_i$ is the degree of the polynomial in $q_i$, $\mathbf{q}$ and $\mathbf{g}$ are vectors of basis functions and parameters respectively, each being of size $P=\prod_{i=1}^{N}P_i$ 
\begin{equation}
\begin{array}{lcl}
	\mathbf{q} = \left(1,\ q_1,\ q_1^2,\ \dots,\ q_2,\ q_1q_2,\ q_1^2q_2,\ \dots \right)^\intercal \\
	\mathbf{g} = \left(g_{00\dots},\ g_{10\dots},\ g_{20\dots},\ \dots ,\ g_{01\dots},\ g_{11\dots},\ g_{21\dots},\ \dots \right)^\intercal
\end{array}
\label{poly_general_basis}
\end{equation}
Although the polynomial model is more universal, it usually contains a much larger number of unknown parameters\footnote{Contrary to the model (\ref{poly_general}), one might consider a model which at first glance appears more suitable, where the force is in the form of a product of single variable polynomials $\tilde F_{i}(q_1,\dots,q_N) = \prod_{m=1}^N P_m(q_m)$. While this model has a much smaller total number of parameters to determine, upon insertion into (\ref{errorf}), we encounter two principal difficulties: (i) if we explicitly perform multiplication and then take the Fourier transform, we obtain a system for the unknown parameters which is nonlinear in the parameters; (ii) if we insert it as it is, the deconvolution problem must be solved $\hat F_i=\hat P_1\ast\dots\ast\hat P_N$ which requires knowledge of the spectral components outside narrow bands surrounding resonances.} which do not directly correspond to physical properties of the material or surface.

Inserting (\ref{poly_general}) into equation (\ref{hat_f}), we obtain a system of linear equations for the polynomial coefficients $g_{i_1\dots i_{N}}$ which is conveniently represented in matrix notation 
\begin{equation}
	\mathbf{g} = \mathbf{\hat{H}}^{+} \mathbf{\hat{F}}_{i} 
%= \mathbf{\hat{H}}^{+} \chi_{i}^{-1} \Delta_h\hat{\mathbf{q}}_{i}
\label{gFind}
\end{equation}
where $\mathbf{\hat{H}}$ is a $B\times P$ matrix with rows $\hat H_{k}=\mathcal{F}_k[\mathbf{q}^\intercal]$ ($k$ is the corresponding component of the discrete Fourier transform of $\mathbf{q}$), $\mathbf{\hat{H}}^{+}$ its Moore-Penrose pseudoinverse, and $\mathbf{\hat{F}}_{i}$ a vector of size $B$\footnote{Strictly speaking, the size of the system is $2B_{i}$ as the Fourier transform of a real function is symmetrical with respect to the zero frequency but complex conjugated. However, this fact does not provide any additional information and can be used only for improving numerical stability of calculations. Solving this system separately for real and imaginary parts gives the same value of $\mathbf{g}$.}. Note that total number of unknown parameters $P$ can not be greater than $B$ and in the case of an overdetermined system ($P<B$) the pseudoinverse will give a unique solution to (\ref{poly_general}) in a least square sense (\ref{errorf}).  Reconstruction of the velocity-dependent part of the force $F_{i}$ can be performed in the same way as for a conservative force, by polynomial expansion in $\dot q_{i}$ with corresponding coefficients.

%-------------------------------------------------------------------------------
\section{\label{sec:results}Reconstruction of two-dimensional tip-surface forces from intermodulation AFM spectra}

Thus the reconstruction of tip-surface force from multiple eigenmodes is a straight-forward generalization of the single eigenmode problem, albeit with the complication of keeping track of multiple modes and the possibility that tones can intermodulate between these modes.  From an algebraic point of view the spectral fitting method can be regarded as a multivariate interpolation \cite{interpol1} and the simplest model which is linear in the parameters suffers from Runge's phenomena \cite{interpol2,interpol3} when high-order nonlinearities couple the multiple eigenmode coordinates.  Furthermore, reconstruction from many eigenmodes is a multi-dimensional problem with many model parameters and it will ultimately suffer from the need to either calibrate or determine these parameters from the limited number of intermodulation products that can be extracted from the narrow frequency bands near the resonances.  

In a real experiment, the AFM detector is typically not able measure over a frequency range which covers many eigenmodes of the cantilever.  Furthermore, the AFM detector is only capable of measuring two signals, corresponding to two orthogonal motions of the cantilever, that of flexing and twisting.  The case of only two eigenmodes is therefore a reasonable simplification of the multi-modal problem which of great practical interest.  In the following we restrict ourselves to this bimodal case.  

We are interested in reconstruction of the two-dimensional tip-surface vector force field $\mathbf{F}_{ts}(\mathbf{r},\dot{\mathbf{r}})$ which depends on the physical tip position in the $y-z$ plane, $\mathbf{r}=(z,y)^\intercal$. Before we proceed with the two-mode analysis we should map the set of forces $F_{i}$ acting on the cantilever eigenmodes $q_i$ onto the physical force $\mathbf{F}_{ts}$. With this aim, we can transform the basis set for defining $q_i$ to separate "pure" modes from "mixed" modes. Applying some transformation, the exact form of which depends on the geometrical shape of the cantilever, we obtain the pure eigencoordinates $z_i$ and $y_i$ contributing only to the tip position perpendicular and parallel to the surface respectively. Here the two different $z_i$ (or $y_i$) are collinear while $z_i$ and $y_i$ are orthogonal. The remaining $\mathbf{q}^\times_i$ are mixed eigencoordinates of the cantilever, or cross-modes contributing to both coordinates $z$ and $y$ simultaneously, so that
\begin{equation}
\begin{array}{lcl}
	z = \sum z_i + \sum\mathbf{q}^\times_i \cdot \mathbf{z}
	\\
	y = \sum y_i + \sum\mathbf{q}^\times_i \cdot \mathbf{y}
\end{array}
	\label{coord_projection}
\end{equation}
Here the second terms are projections of the mixed eigencoordinates onto the tip coordinate system $(\mathbf{z},\mathbf{y})$. In doing so, the force $\mathbf{F}_{ts}$ is projected onto the $(\mathbf{z},\mathbf{y})$ so that its components parallel to the surface $F_z$ and perpendicular to the surface $F_y$ act on the corresponding pure modes, i.e. $F_i=F_z$ for each $z_i$, $F_i=F_y$ for each $y_i$, and different force projections act on the mixed modes.

Simultaneous excitation of the several pure eigenmodes coupled by nonlinear tip-surface interaction leads to excitation of the mixed modes which allows for measurements of the response in additional frequency bands. Although it would provide additional information, in this paper we investigate the simplest multimodal motion, that of pure bimodal motion without considering the cross-modes. In this case, analysis of the cantilever dynamics reduces to study of two characteristic regimes: bimodal motion of the collinear eigencoordinates (e.g. two flexural or torsional modes) and orthogonal eigencoordinates (one flexural and one torsional mode). Let us proceed with the former.

%-------------------------------------------------------------------------------
\subsection{\label{subsec:case1}Case 1: Two collinear modes}
This case corresponds to the dynamics of two flexural modes $z_1$ and $z_2$, so that the total perpendicular tip deflection $z=z_1+z_2$
\begin{equation}
\begin{array}{lcl}
	k_1\left(\frac{1}{\omega_1^2}\ddot z_1 + \frac{1}{Q_1\omega_1}\dot z_1 + z_1 \right)= F_z(z,\dot z) + \mathrm{f}_{1}(t)
\\
	k_2\left(\frac{1}{\omega_2^2}\ddot z_2 + \frac{1}{Q_2\omega_2}\dot z_2 + z_2 \right)= F_z(z,\dot z) + \mathrm{f}_{2}(t)
\end{array}
	\label{case1_worksys}
\end{equation}
Here, on the right hand side we have the same vector component $F_z$ of the tip-surface force field depending on $z$ and $\dot z$.

\begin{figure}
\centering
\includegraphics[width=75mm]{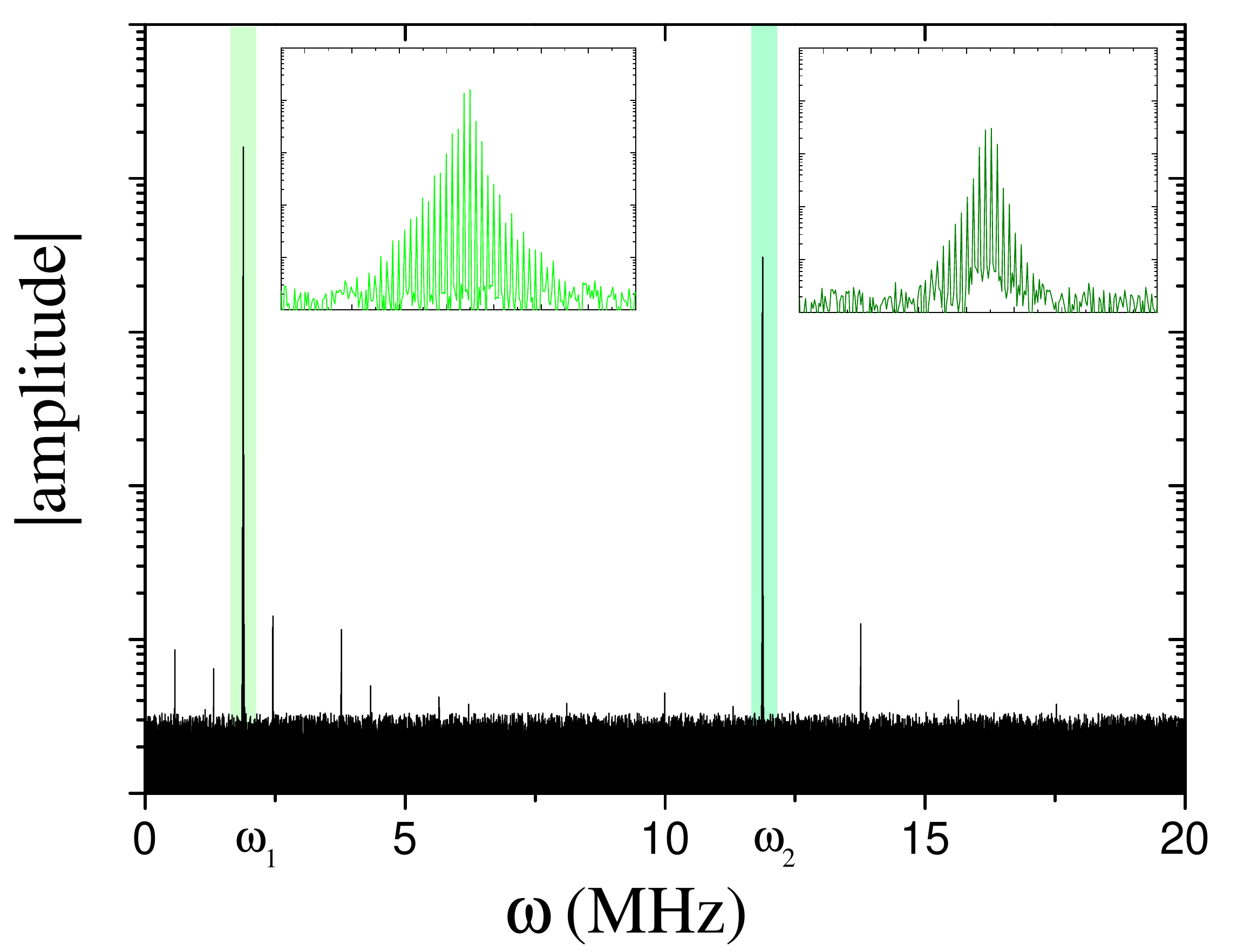}
 \caption{\label{fig:imspectr}(Color online) Spectrum of the engaged cantilever motion with two flexural modes. Two highest peaks are clearly seen above noise level near resonant frequencies $\omega_1$ and $\omega_2$ which consist of the intermodulation products depicted in insertions. The spectrum is obtained by integrating the system (\ref{case1_worksys}) with the tip-surface force (\ref{case1_vdW-DMT}) and consequent addition of a white noise.}
\end{figure}

\begin{figure}
\centering
\includegraphics[width=75mm,angle=0]{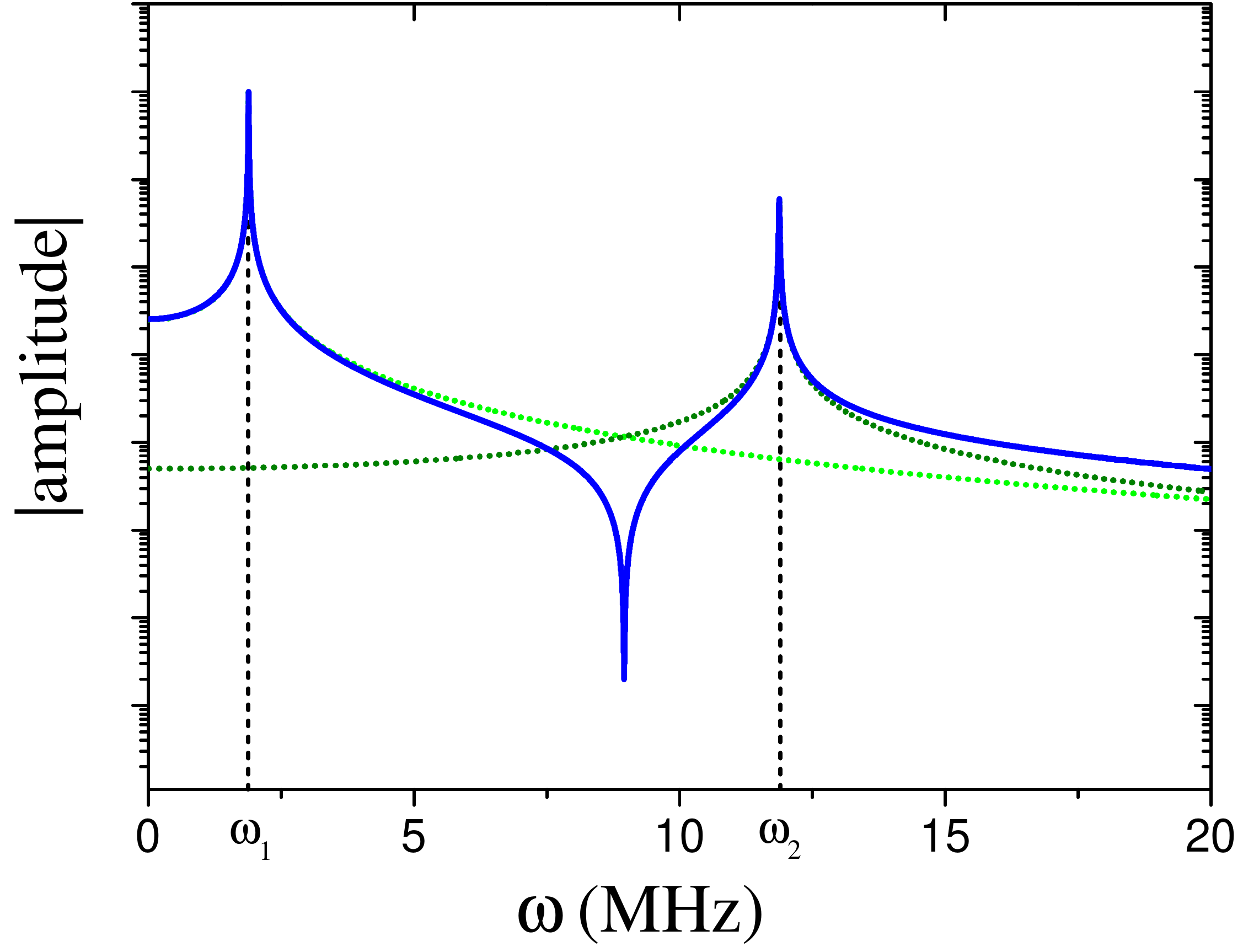}
 \caption{\label{fig:transfunct}(Color online) Total transfer function $\chi$ (blue solid line) for the cantilever with two flexural modes with transfer functions $\chi_1$ and $\chi_2$ (light and dark green dotted lines respectively).}
\end{figure}

Firstly, let us consider a model for the position-dependent part of $F_z$ in some general form
\begin{equation}
	\tilde F_z(z_1,z_2) = \sum\limits_{i=0}^{P_{z_1}}\sum\limits_{j=0}^{P_{z_2}} g_{ij} z_1^i z_2^j
\label{case1_model1}
\end{equation}
This model requires determination of a large number $P_z(P_z-1)/2$ of coefficients $g_{ij}$ in order to define the polynomial of order $P_z=P_{z_1}P_{z_2}$. If we use the fact that $F_z(z_1,z_2)=F_z(z_1+z_2)$ we can define a polynomial of the form 
\begin{equation}
	\tilde F_z(z_1+z_2) = \sum\limits_{i=0}^{P_z}g_i (z_1+z_2)^i
\label{case1_model2}
\end{equation}
which contains only $P_z+1$ unknown coefficients $g_i$. In this case we should consider not two separate spectra of dynamic variables $\hat z_1$ and $\hat z_2$ but one united spectrum $\hat z=\hat z_1 + \hat z_2$ which is actually measured. As a result, it is possible to obtain a tip-surface interaction spectrum $\hat F_z$ (fig.~\ref{fig:imspectr})
\begin{equation}
	\hat F_z = \chi^{-1}\Delta_h \hat z
\label{case1_imsZ}
\end{equation}
making use of the total response function\footnote{In principle, it can be generalized for case of $N$ collinear modes $\chi = \sum_{i=0}^{N}\chi_i$.} $\chi = \chi_1+\chi_2$ depicted in fig.~\ref{fig:transfunct}.

\begin{figure*}
\centering
\includegraphics[width=75mm]{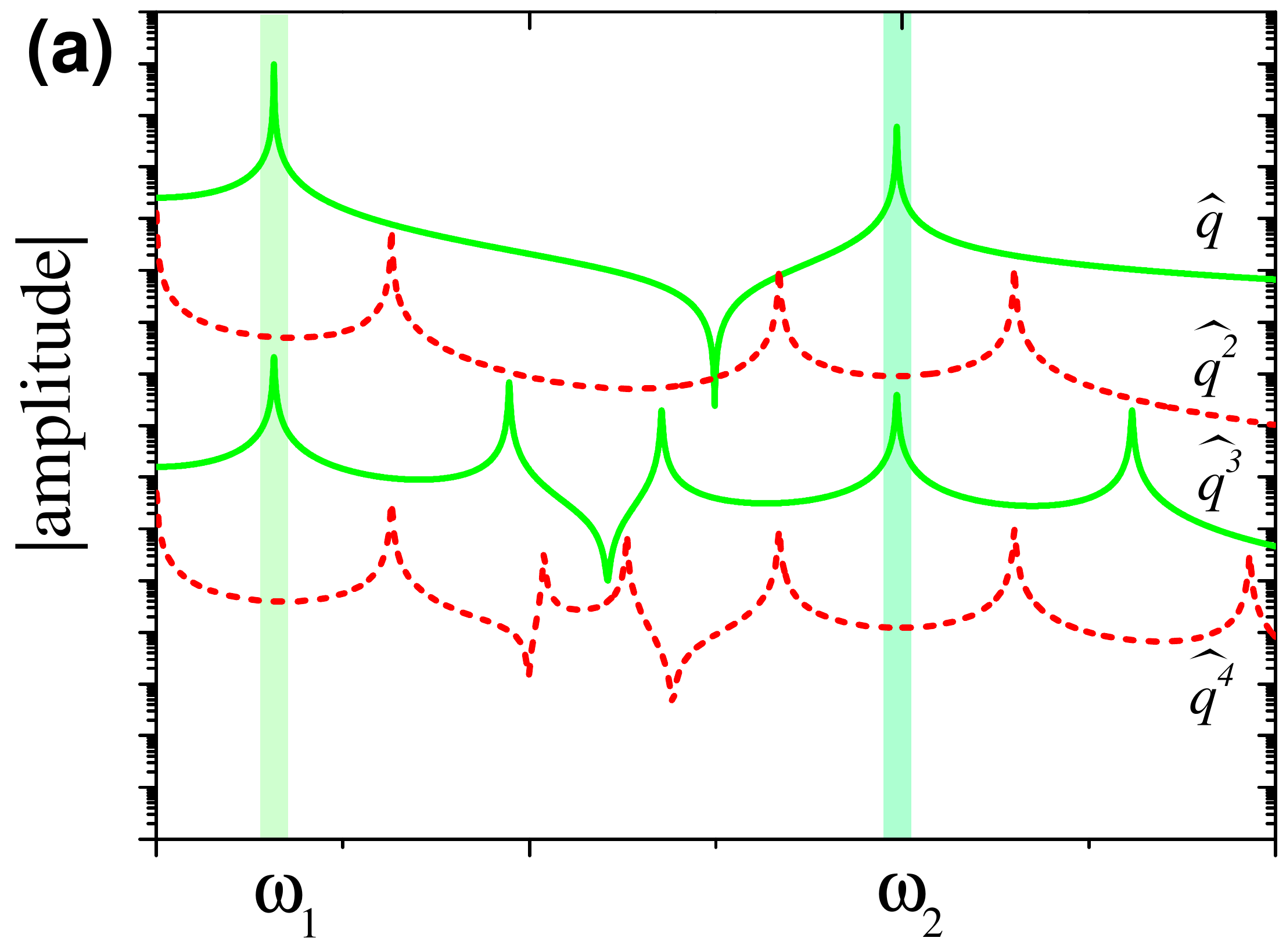}\hspace{5mm}
\includegraphics[width=75mm]{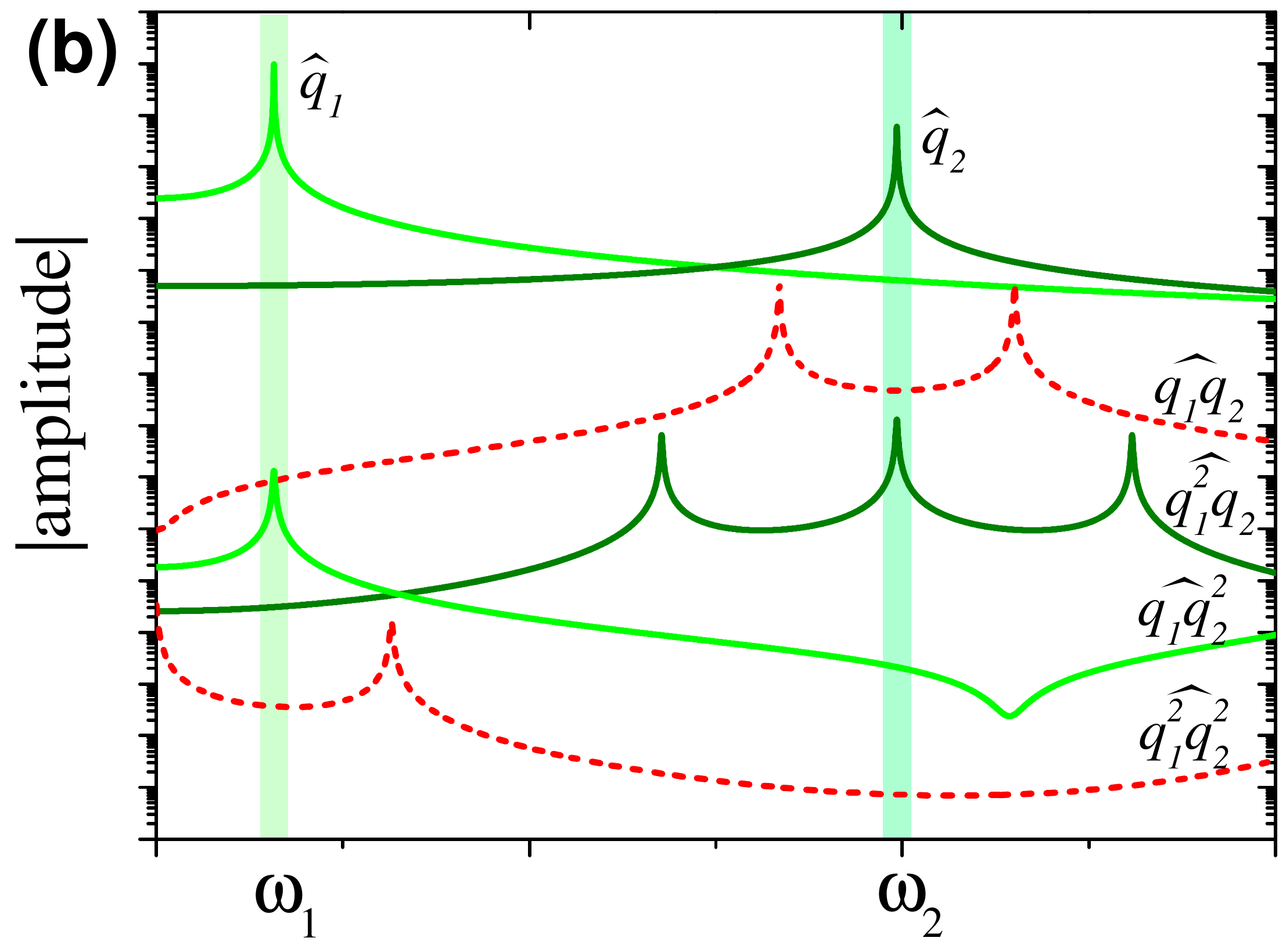}\\
 \caption{\label{fig:zPowers}(Color online) (a) Response spectra for powers of single dynamic variable with two resonances and (b) two separate variables with one resonance. Two frequency windows near resonances $\omega_1$ and $\omega_2$ are highlighted with light and dark green colors respectively. Red dashed line means that the corresponding maximum response lies outside the frequency bands.}
\end{figure*}

Prior to force reconstruction it is necessary to investigate what kind of information about $F_z$ is contained in the spectral bands for $z_1$ and $z_2$. In the current investigation we try to reconstruct the force using information contained only in the narrow frequency bands near $\omega_1$ and $\omega_2$, so all weak response peaks outside these bands are discarded. We also require that the second resonance frequency $\omega_2$ is not an integer multiplier of the $\omega_1$, so the second band does not capture any higher harmonics and intermodulation products produced by the drive in the first band. If $\omega_2$ were a harmonic of $\omega_1$ it would be of considerable advantage for force measurement \cite{harm_cant1}. Considering the monomial basis (\ref{case1_model2}) we can approximately evaluate the Fourier spectrum of the $i^\mathrm{th}$ power $\mathcal{F}[z^k] = \mathcal{F}[z]\ast\dots\ast\mathcal{F}[z]$ via convolution of the transfer function $\chi$ with itself, which gives an upper bound of the response in the frequency domain. Figure \ref{fig:zPowers}(a) demonstrates that only components with odd powers $i$ have significant value in the narrow bands near the resonances. Consequentially, only parameters $g_i$ for odd powers $z^i$ ($i$ odd) can be found directly using the measured spectrum.

Having solved the system for the odd parameters, we can use them to recover the even parameters by applying an additional constraint: that the tip-surface force for tip positions above its rest point equals zero $F_z(z>0)=0$ which leads to the system
\begin{equation}
	\sum\limits_{i=0}^{N/2}g_{2i}z_k^{2i}=\sum\limits_{i=0}^{N/2}g_{2i+1}z_k^{2i+1}
\label{case1_syseven}
\end{equation}
for all $z_k>0$ measured at discrete time moments $t_k$.

Reconstruction of the velocity-dependent part of $F_z$ is achieved by adding a new variable $\dot z=\dot z_1+\dot z_2$ to the model (\ref{case1_model2}) which consequently increases the number of parameters by the degree of the polynomial in $\dot z$
\begin{equation}
	\tilde F_z(z,\dot z) = \sum\limits_{i=0}^{P_z}\sum\limits_{j=0}^{P_{\dot z}} g_{ij} z^i\dot z^j
\label{case1_model3}
\end{equation}
Noticing that $\mathcal{F}[z^i\dot z^j] \propto \mathcal{F}[z^{i+j}]$, only coefficients $g_{ij}$ in front of $z^i\dot z^j$ where $i+j$ is odd can be determined from the measured spectrum. While coefficients before $z\dot z$, $z^3\dot z$, $z\dot z^3$, etc. can be found using the system (\ref{case1_syseven}).

\begin{figure*}
\centering
\includegraphics[width=75mm]{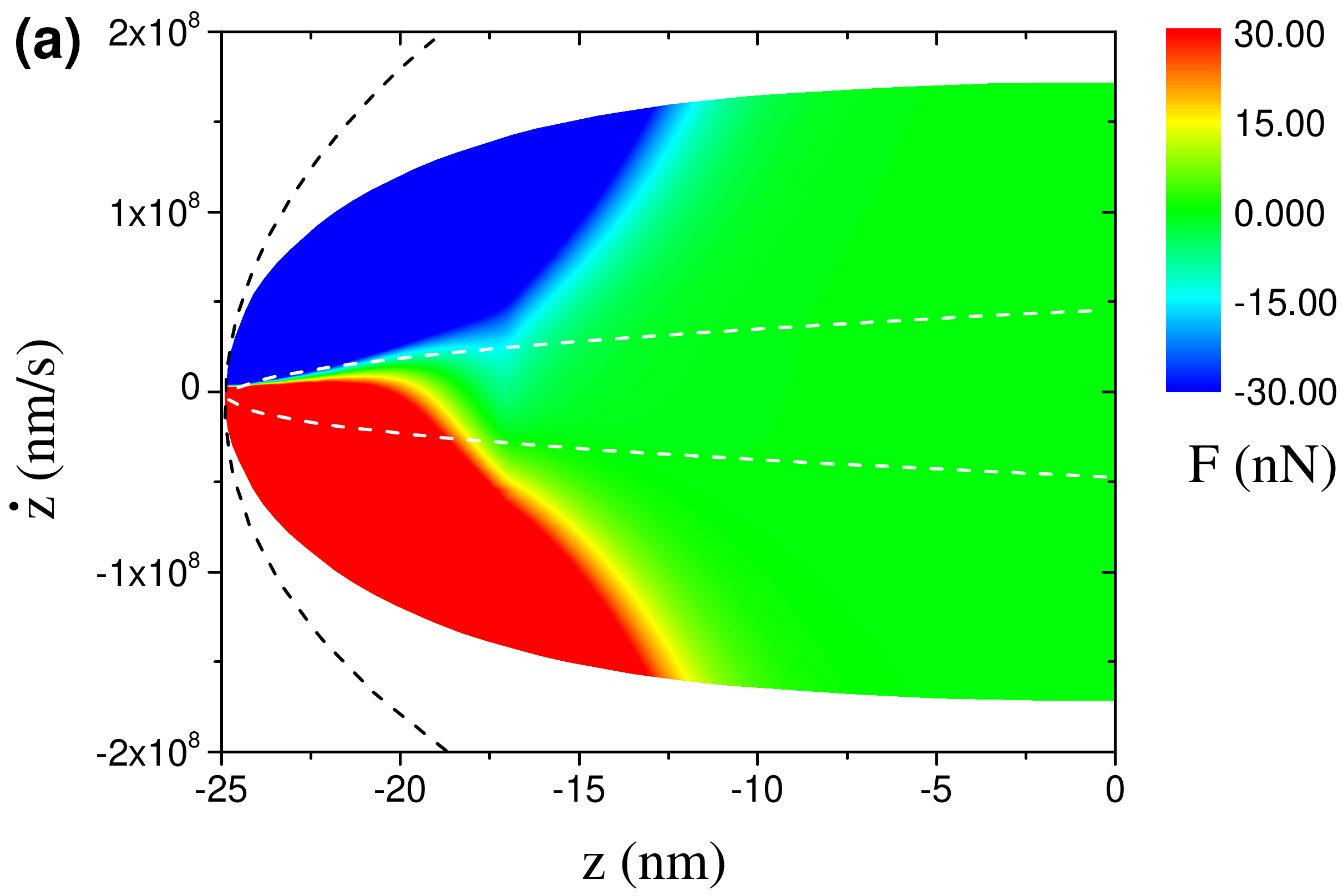}\hspace{5mm}
\includegraphics[width=77mm]{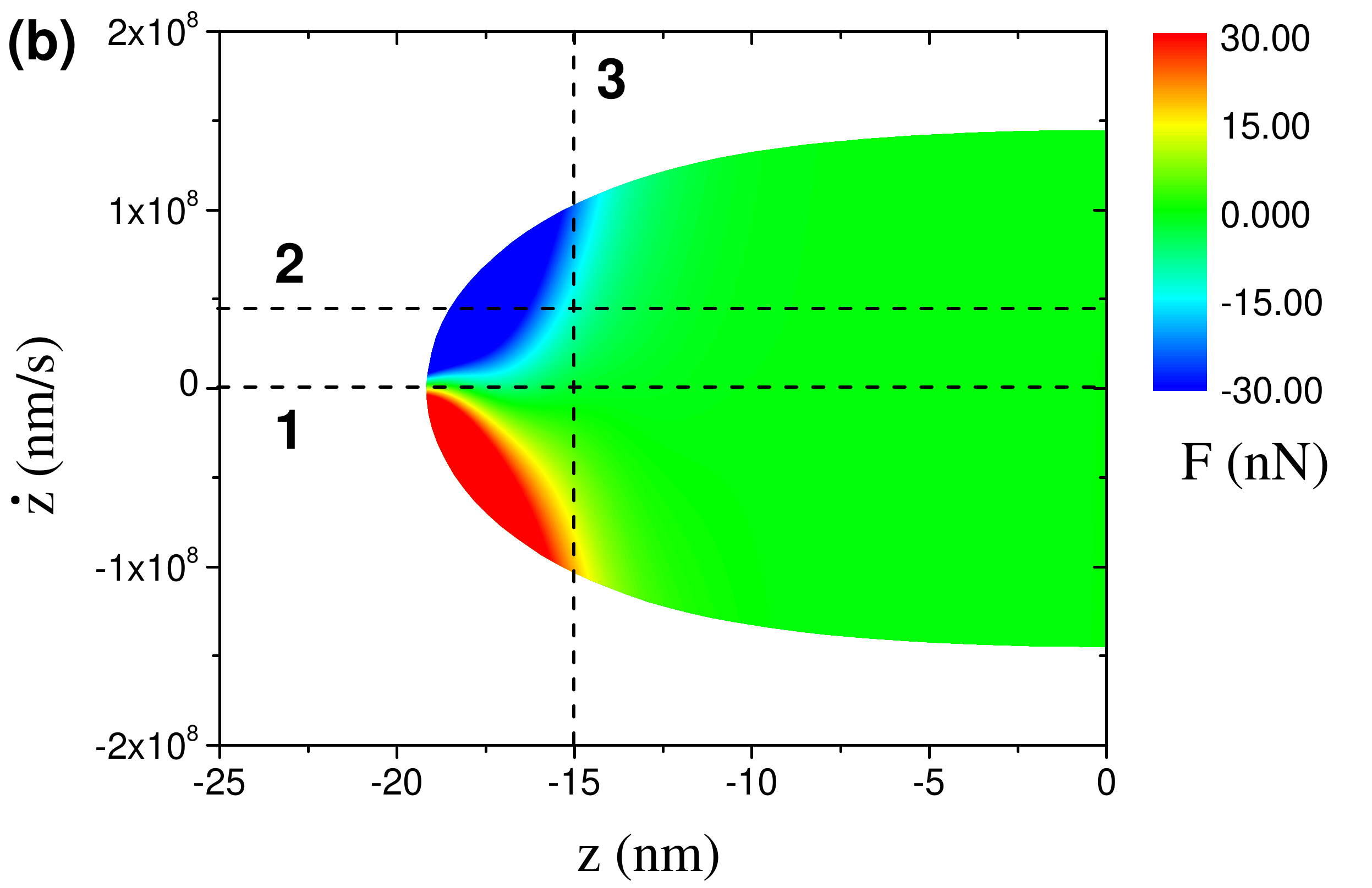}\\
 \caption{\label{fig:1D_f_diss}(Color online) (a) vdW-DMT force with position dependent damping in the region of the free tip motion with bimodal drive. White and black dashed lines confine covered domains of phase space in case of single mode drive of the first and second modes respectively (difference between maximum velocities is about order of magnitude). In all three cases, total maximum amplitude of oscillations is kept constant and equals $25$ nm. (b) Reconstructed force in the region of the engaged tip motion. Its cross-sections 1--3 are depicted in fig.~\ref{fig:1D_f_diss_cuts}(a)--(c) to highlight agreement between the actual force used in simulation and the reconstructed force.}
\end{figure*}

We simulate the collinear bimodal case using the CVODE integrator \cite{CVODE} with: $\omega_1 = 2\pi\,300$ kHz, $k_1 = 40$ N/m, $Q_1 = 400$, $\omega_2 = 6.3\omega_1$, $k_2 = 50k_1$, $Q_2 = 3Q_1$ (ratios for the second mode are taken from \cite{calib1}). The driving forces $\mathrm{f}_{1,2}$ are chosen to have the same phase and give equal maximum free response (when $F_z\equiv 0$) at each mode $A_{z_1}=A_{z_2}=12.5$ nm so the total maximum amplitude of oscillations is $A_z=25$ nm; all four drive frequencies $\omega_{1,2}\pm\delta\omega/2$ are integer multipliers of base frequency $\delta\omega=2\pi\,0.2$ kHz. The engaged height $h$ above the surface is $17$ nm. The model of the tip-surface force $F_z$ is the vdW-DMT force \cite{vdW-DMT} with the nonlinear damping term exponentially dependent on the tip position 
\begin{equation}
\begin{array}{lcl}
	F_z(z,\dot z) = F_z^{con}(z) + F_z^{dis}(z,\dot z)\\
	F_z^{con}(z) = \begin{cases} 
		-\frac{HR}{6z^2}, &z\geq a_0 \\
		-\frac{HR}{6a_0^2}+\frac{4}{3}E^{*}\sqrt{R(a_0-z)}, &z<a_0
	\end{cases}\\
	F_z^{dis}(z,\dot z) = - \left(\gamma_1\dot z + \gamma_3\dot z^3\right)e^{-z/\lambda_z}
\end{array}
\label{case1_vdW-DMT}
\end{equation}
with the following seven phenomenological parameters: intermolecular distance $a_0=0.3$ nm, Hamaker constant $H=7.1\times 10^{-20}$ J, effective modulus $E^{*}=1.0$ GPa, tip radius $R=10$ nm, damping decay length $\lambda_z=1.5$ nm, damping factors $\gamma_1=2.2\times 10^{-7}$ kg/s and $\gamma_3 = 10^{-22}$ kg$\cdot$s/m$^2$ (fig.~\ref{fig:1D_f_diss}(a)). 
It is worth noting that the calibration of the higher eigenmodes parameters $\omega_i$, $Q_i$ and $k_i$ which is required for force reconstruction, is itself a challenging task in multimodal AFM \cite{calib1,calib2,calib3}.

Using $B_{1,2}=24$ peaks in each band (fig.~\ref{fig:imspectr}) for the reconstruction, we assume the model (\ref{case1_model3}) degree in $z$ to be $P_z=21$. Numerical results show that the force reconstruction using higher powers of $\dot z^{i}$ ($i>1$) is less reliable as it encounters difficulties connected with multivariate interpolation. Although two modes give us twice the number of spectral components in comparison with singlemode case, this additional information is still insufficient for reconstruction of nonlinear (non-viscous) damping. Therefore, we are restricted the model to be linear in $\dot z$. 

\begin{figure*}
\centering
\includegraphics[width=55mm]{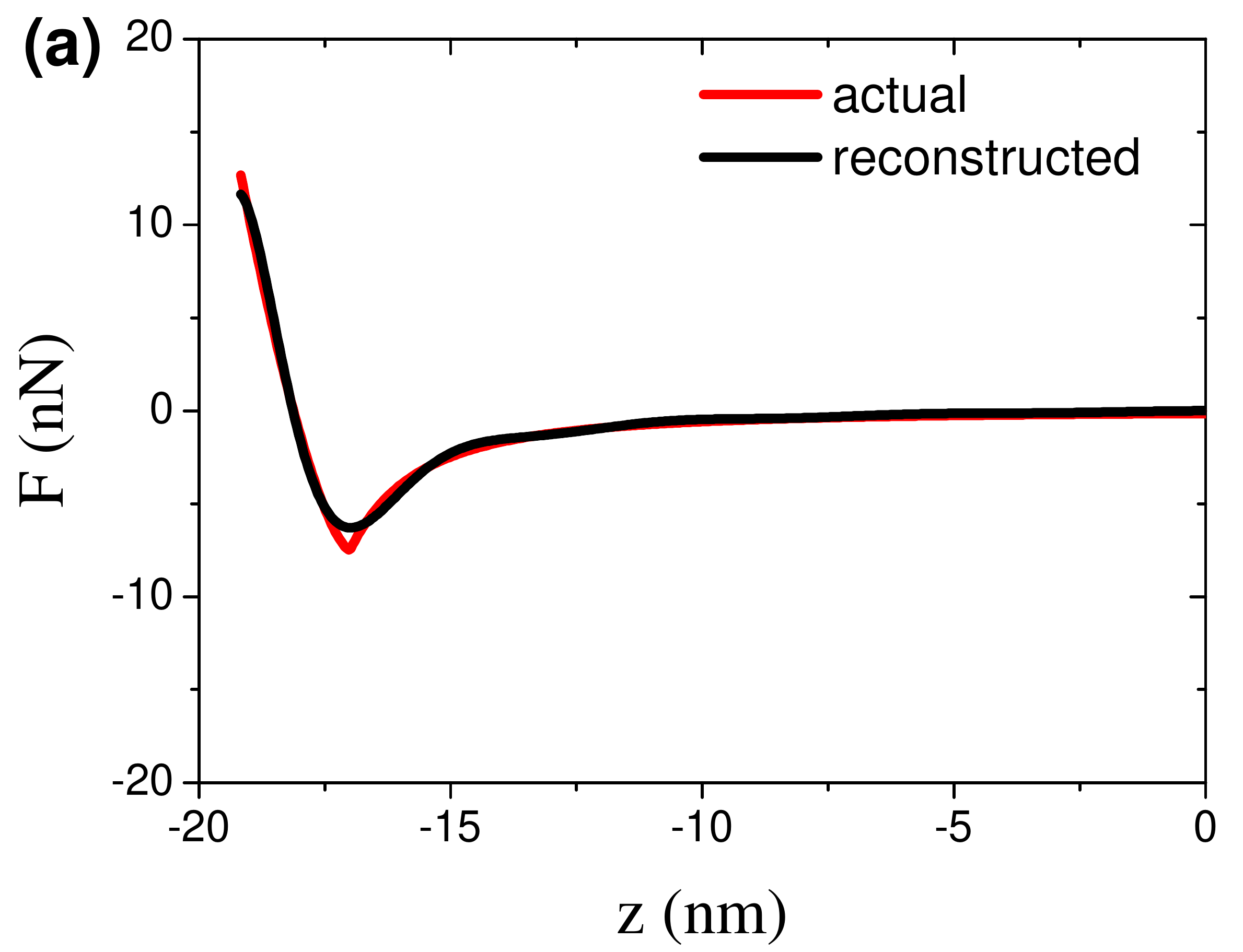}
\includegraphics[width=56mm]{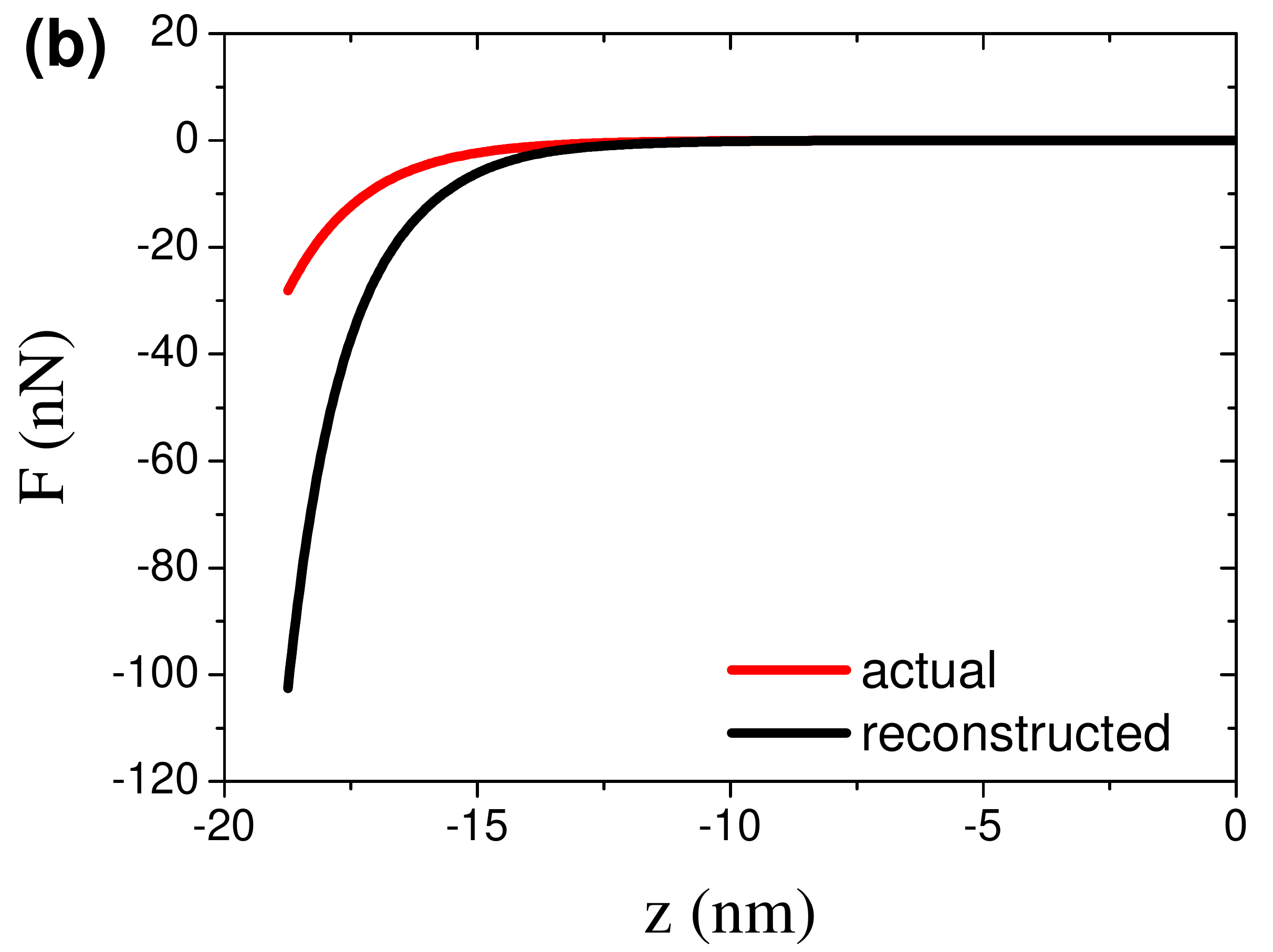}
\includegraphics[width=57mm]{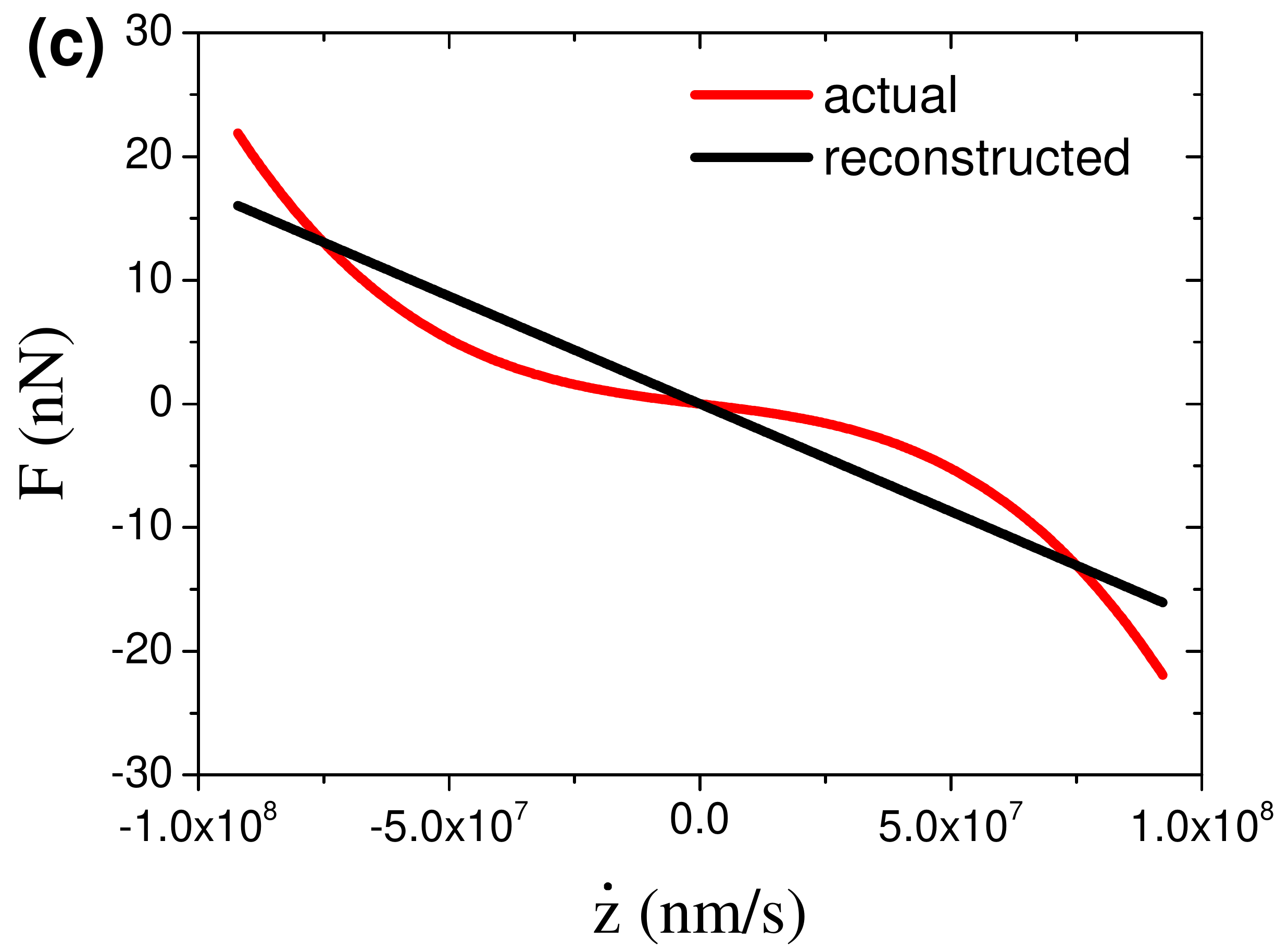}\\
 \caption{\label{fig:1D_f_diss_cuts}(Color online) Cross-sections of the reconstructed tip-surface force (black) with comparison to the actual force used in simulation (red) for the cantilever driven at two flexural modes. (a) $\tilde F_z(z,\dot z = 0)$; (b) $\tilde F_z(z,\dot z = 0.2\dot z_{\max})-\tilde F_z(z,\dot z = 0)$; (c) $\tilde F_z(z=0.75z_{\min},\dot z)$. Linear fit for nonlinear damping catches nicely overall trend and conservative part demonstrates an excellent agreement with the actual force.}
\end{figure*}

This numerical analysis suggests that there is no difference in the quality of the reconstructed force using intermodulation products from frequency bands surrounding the first, second or both resonances. We show results for the reconstructed $\tilde F_z(z,\dot z)$ and its cross-sections in fig.~\ref{fig:1D_f_diss}(b) and fig.~\ref{fig:1D_f_diss_cuts} using only the resonant detection band around the first eigenmode. 
The linear fit (\ref{case1_model3}) for the model with nonlinear damping (\ref{case1_vdW-DMT}) nicely captures the overall trend of the dissipative part and the reconstructed conservative part demonstrates excellent agreement with the actual force. If we simulate the system (\ref{case1_worksys}) using the force $F_z$ linear in $\dot z$, for instance assuming $\gamma_3=0$ in (\ref{case1_vdW-DMT}), the reconstructed force $\tilde F_z$ shows nearly perfect agreement with the actual force $F_z$ (figures are not included).

\begin{figure*}
\centering
\includegraphics[width=75mm]{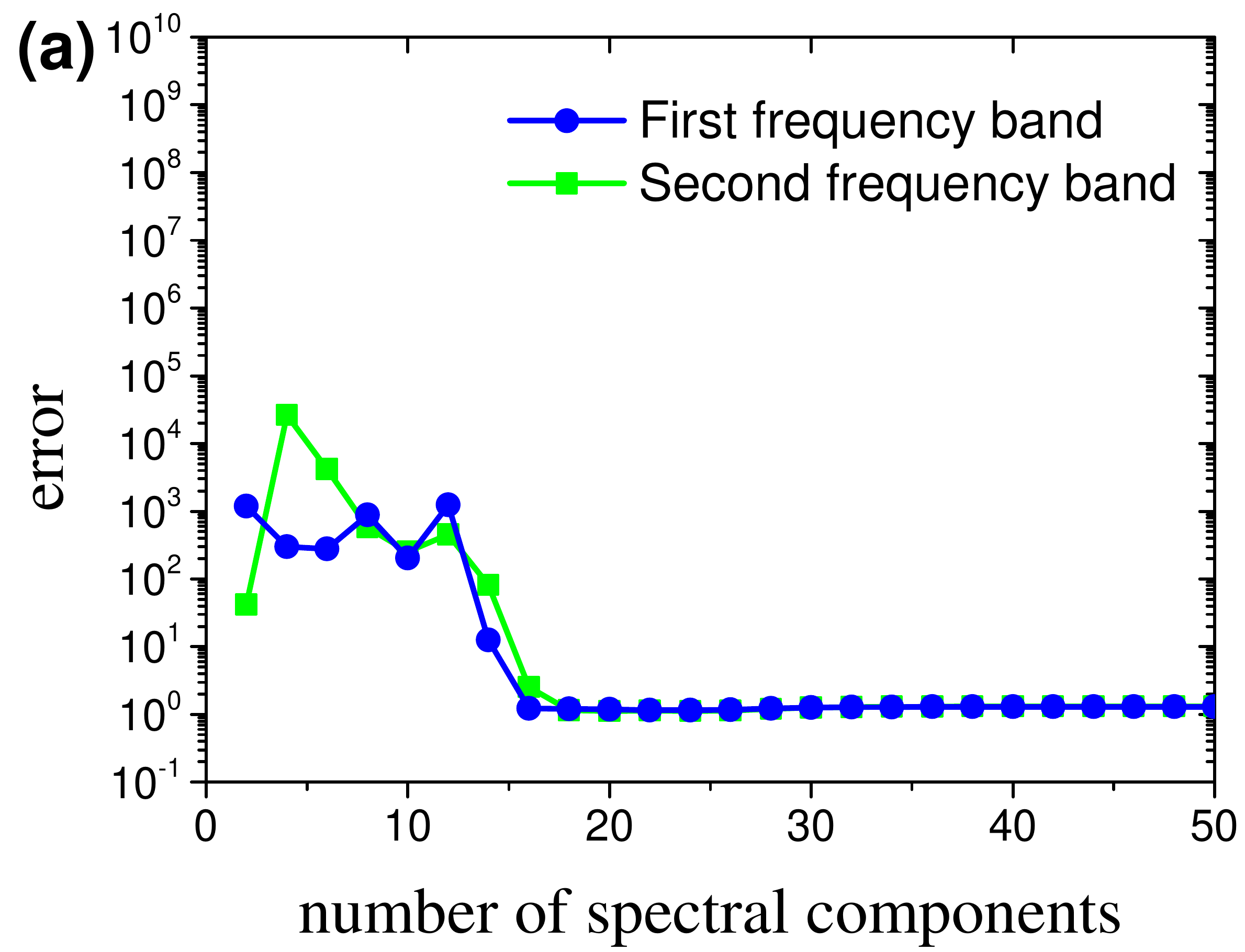}\hspace{5mm}
\includegraphics[width=75mm]{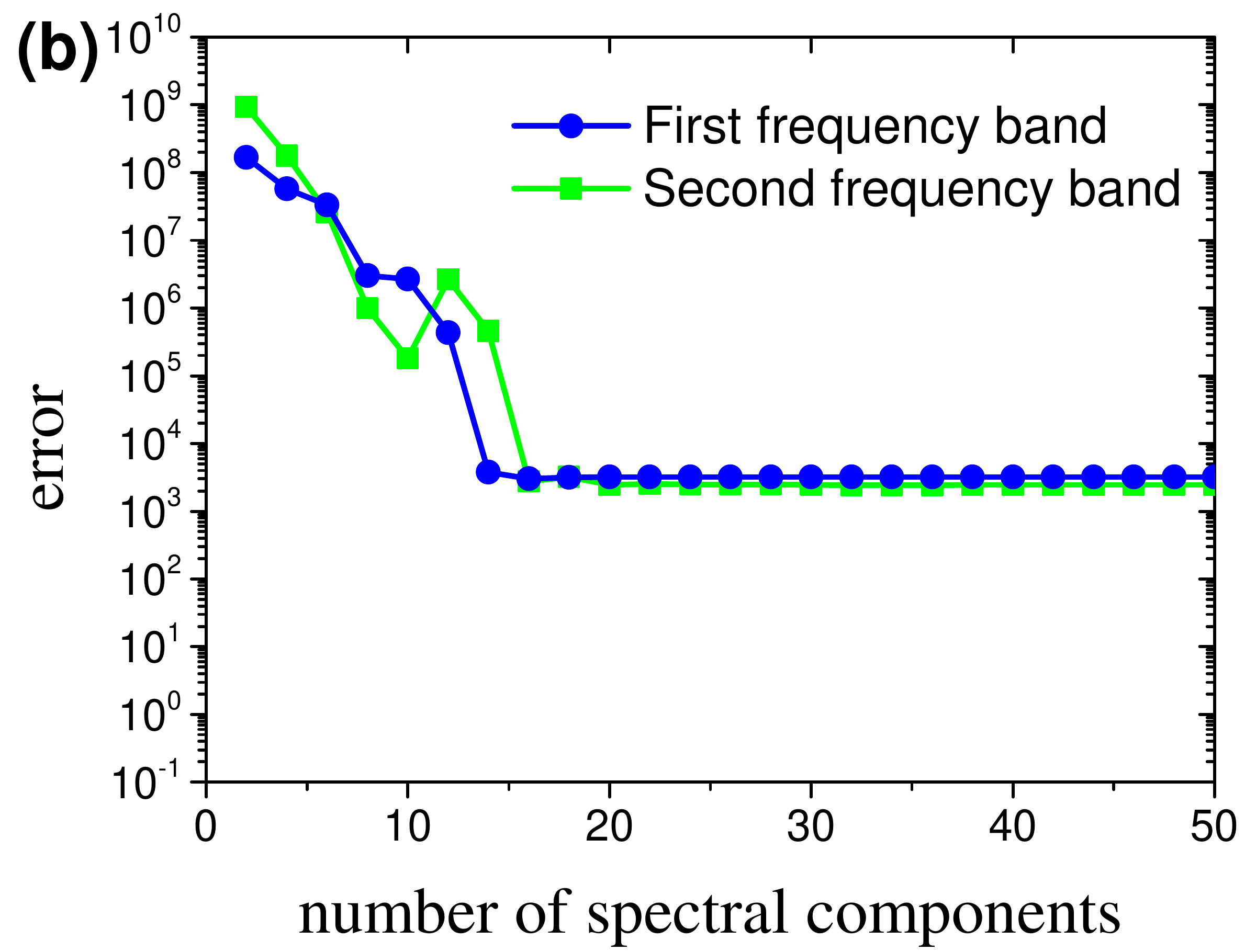}\\
 \caption{\label{fig:errorPeaks}(Color online) Absolute error of reconstruction of (a) conservative part of the tip-surface force (fig. \ref{fig:1D_f_diss_cuts}(a)) and (b) dissipative part (fig. \ref{fig:1D_f_diss_cuts}(b)) versus number of spectral components taken into account in the first frequency band (blue circles) and in the second (green squares).}
\end{figure*}

One can compare the information contained in the two frequency bands near each eigenmode resonance by estimating the quality of the reconstruction as a function of the number of spectral components $B_{1,2}$ used in the reconstruction, as the least square error function
\begin{equation}
	e_i(B_i)=\int\limits_{z_{\min}}^{z_{\max}} \left[\tilde F_z(z,\dot z|B_i)-F_z(z,\dot z)\right]^2\mathrm{d}z
\label{case1_intError}
\end{equation}
This error function plotted versus the number of spectral components is depicted in fig.~\ref{fig:errorPeaks}.
We see similar behavior for both bands: significant drop for the number of spectral components lager than half of the polynomial power $P_z=21$ in the expansion of $\tilde F_z$, and no qualitative improvements after reaching a number of spectral components less than $P_z$. 

Another interesting observation regards the reconstruction of a dissipative force which turns-on only for the tip velocities $\dot z$, higher than the maximum velocity of the first mode $z_1$ (for some constant amplitude) and is zero otherwise. Information from both bands gives approximately the same reconstructed curves with showing overall trend of the dissipative part and in excellent agreement for the conservative part (plots not shown). However, reconstructing with spectral components from only the first frequency band yields more accurate dissipative force approximation as $z_1$ has smaller stored oscillation energy and, consequently, is more vulnerable to the dissipative force than $z_2$. Nonetheless, if we excite only the first mode, we would not be able to reconstruct this "threshold" dissipative force at all, as the magnitude of the tip velocity would not be enough to turn on the dissipation. Thus, simultaneous excitation of two eigenmodes allows to explore a wider region of the phase space of the tip motion (fig.~\ref{fig:1D_f_diss}(a)) while keeping the total maximum amplitude constant.

%-------------------------------------------------------------------------------
\subsection{\label{subsec:case2}Case 2: Two orthogonal modes}
This case corresponds to the dynamics of one flexural $z$ and one torsional $y$ mode
\begin{equation}
\begin{array}{lcl}
	k_z\left(\frac{1}{\omega_z^2}\ddot z + \frac{1}{Q_z\omega_z}\dot z + z \right) = F_z(z,y) + \mathrm{f}_{z}(t)
\\
	k_y\left(\frac{1}{\omega_y^2}\ddot y + \frac{1}{Q_y\omega_y}\dot y + y \right) = F_y(z,y) + \mathrm{f}_{y}(t)
\end{array}
	\label{case2_worksys}
\end{equation}
with two different projections of the conservative force, $F_z$ and $F_y$ on the r.h.s. 

We start from the polynomial models for reconstruction
\begin{equation}
\begin{array}{lcl}
	\tilde F_z(z,y) = \sum\limits_{i=0}^{P_z}\sum\limits_{j=0}^{P_y} g^{(z)}_{ij} z^i y^j \\
	\tilde F_y(z,y) = \sum\limits_{i=0}^{P_z}\sum\limits_{j=0}^{P_y} g^{(y)}_{ij} z^i y^j
\end{array}
\label{case2_models}
\end{equation}
As for case 1 it is not possible to find all parameters of these models. Firstly, as we are limited in number of measurable intermodulation products and therefore in maximum degree of the polynomial. We choose $z$ direction as the most interesting degree of freedom, by which we mean that the maximum degree of the polynomial in this variable will be much higher than for $y$. In accordance to the fig.~\ref{fig:zPowers}(b), the captured information about forces $\tilde F_z$ will be odd in $z$ and even in $y$ and vice versa for $\tilde F_y$
\begin{equation}
\begin{array}{lcl}
	\tilde F_z(-z,\pm y) = -\tilde F_z(z,y)\\
	\tilde F_y(\pm z,-y) = -\tilde F_y(z,y)
\end{array}
\label{case2_propSign12}
\end{equation}
as the first flexural resonance $\omega_z$ is typically far lower in frequency than the first torsional resonance $\omega_y$ \cite{gen1_cantDyn}. It is possible to recover the coefficients $g^{(z)}_{ij}$ of even powers of $y$ and $g^{(y)}_{ij}$ of odd powers of $y$ (when $i+j$ is even) by using the additional constraint for $z$ dependence of the force components $F_{z,y}(z>0,y)=0$ and eq. (\ref{case1_syseven}). While the information about all coefficients of $\tilde F_z$ with odd powers of $y$ and $\tilde F_y$ with even powers of $y$ is lost because we have no such constraint on the $y$ dependence.

\begin{figure*}
\centering
\includegraphics[width=75mm]{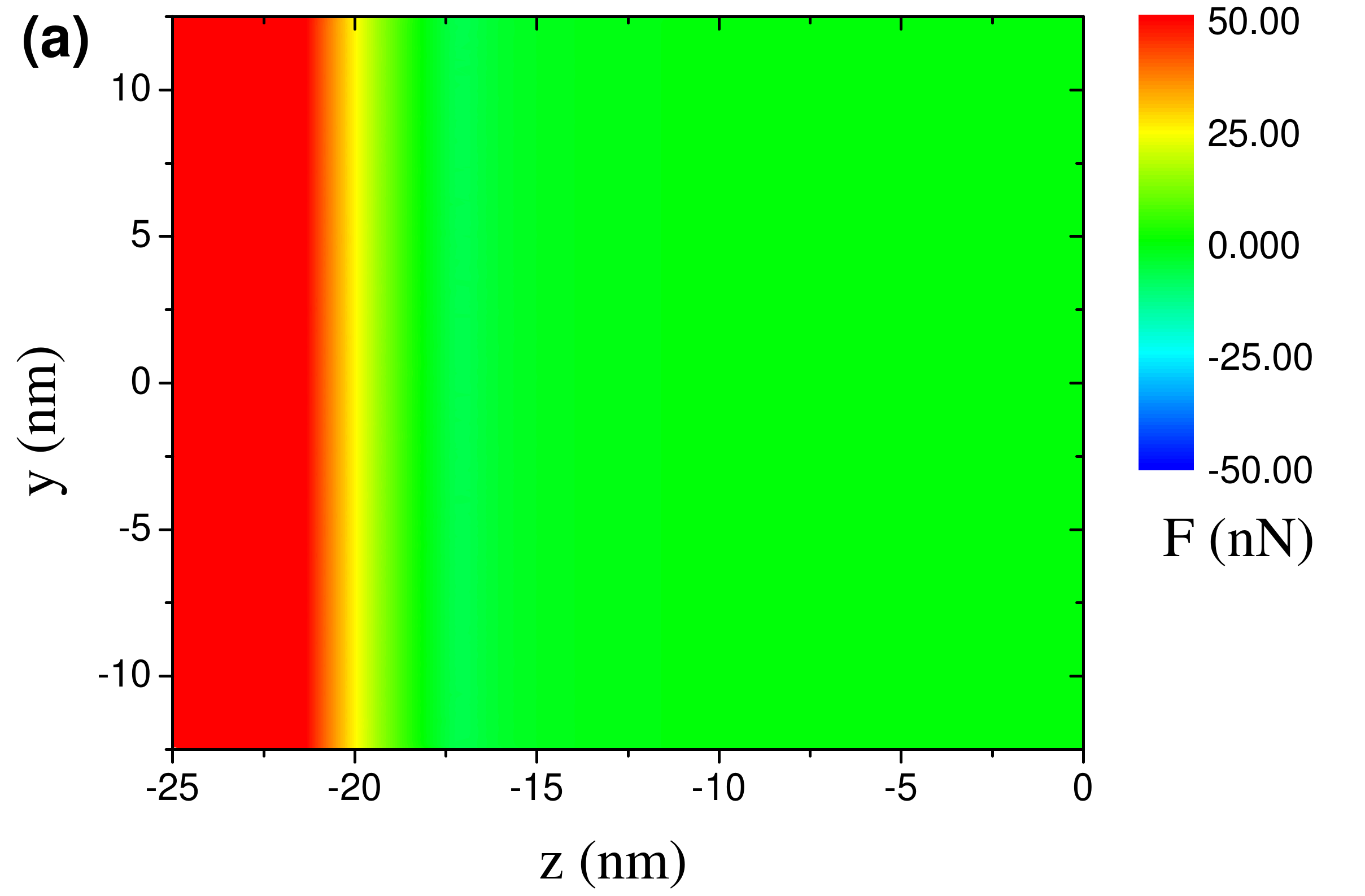}\hspace{5mm}
\includegraphics[width=75mm]{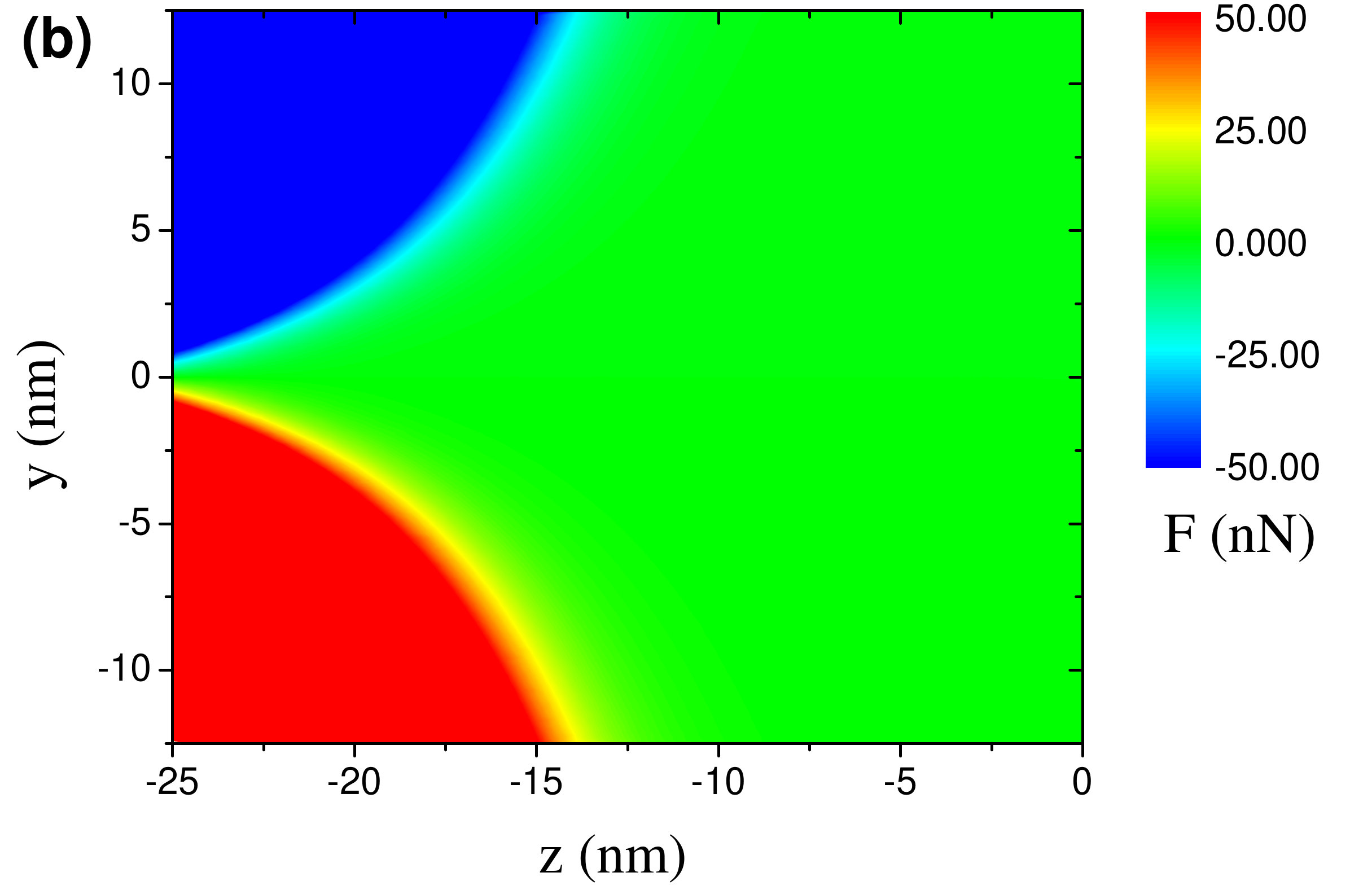}\\
\vspace{5mm}
\includegraphics[width=75mm]{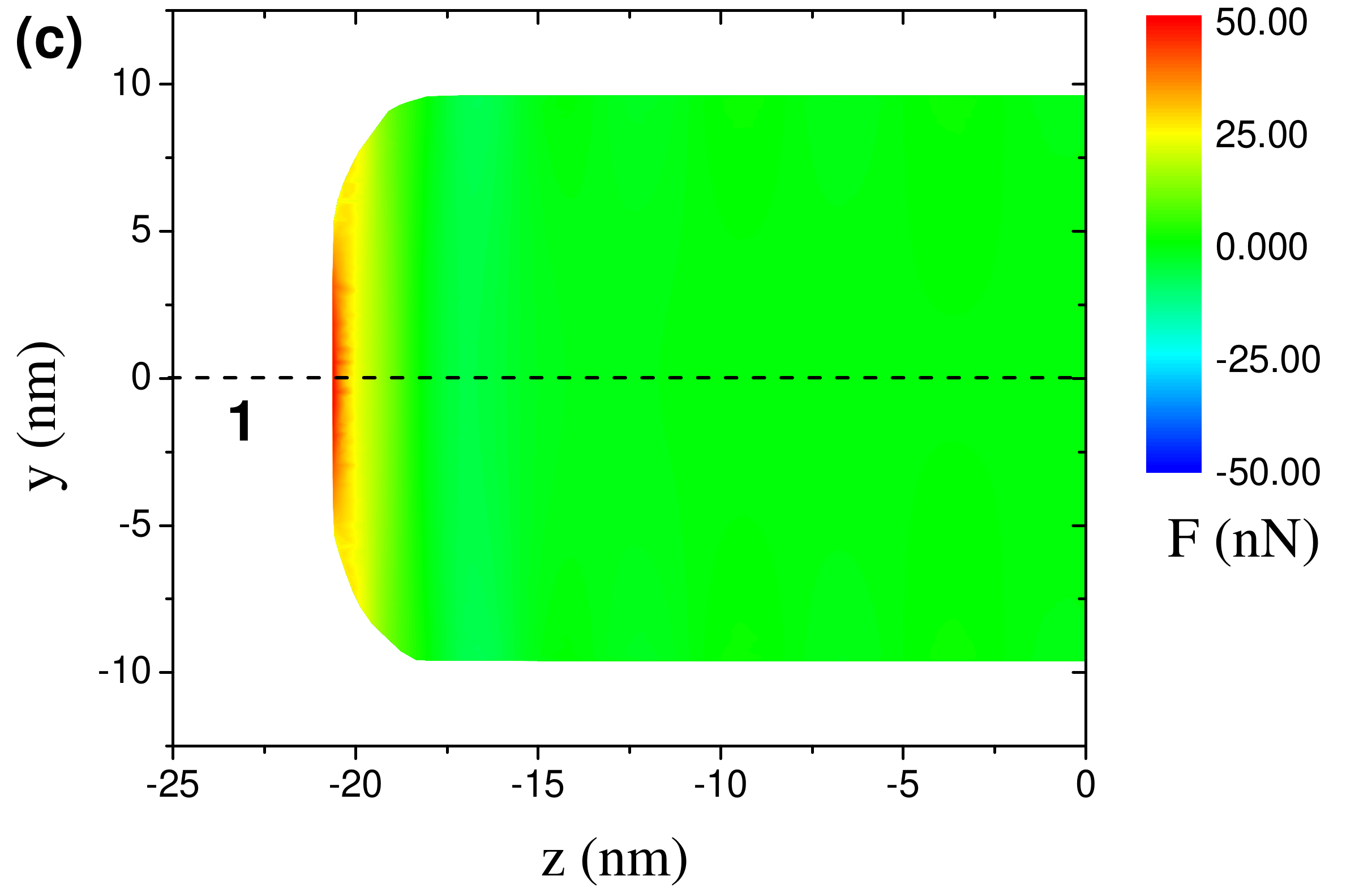}\hspace{5mm}
\includegraphics[width=75mm]{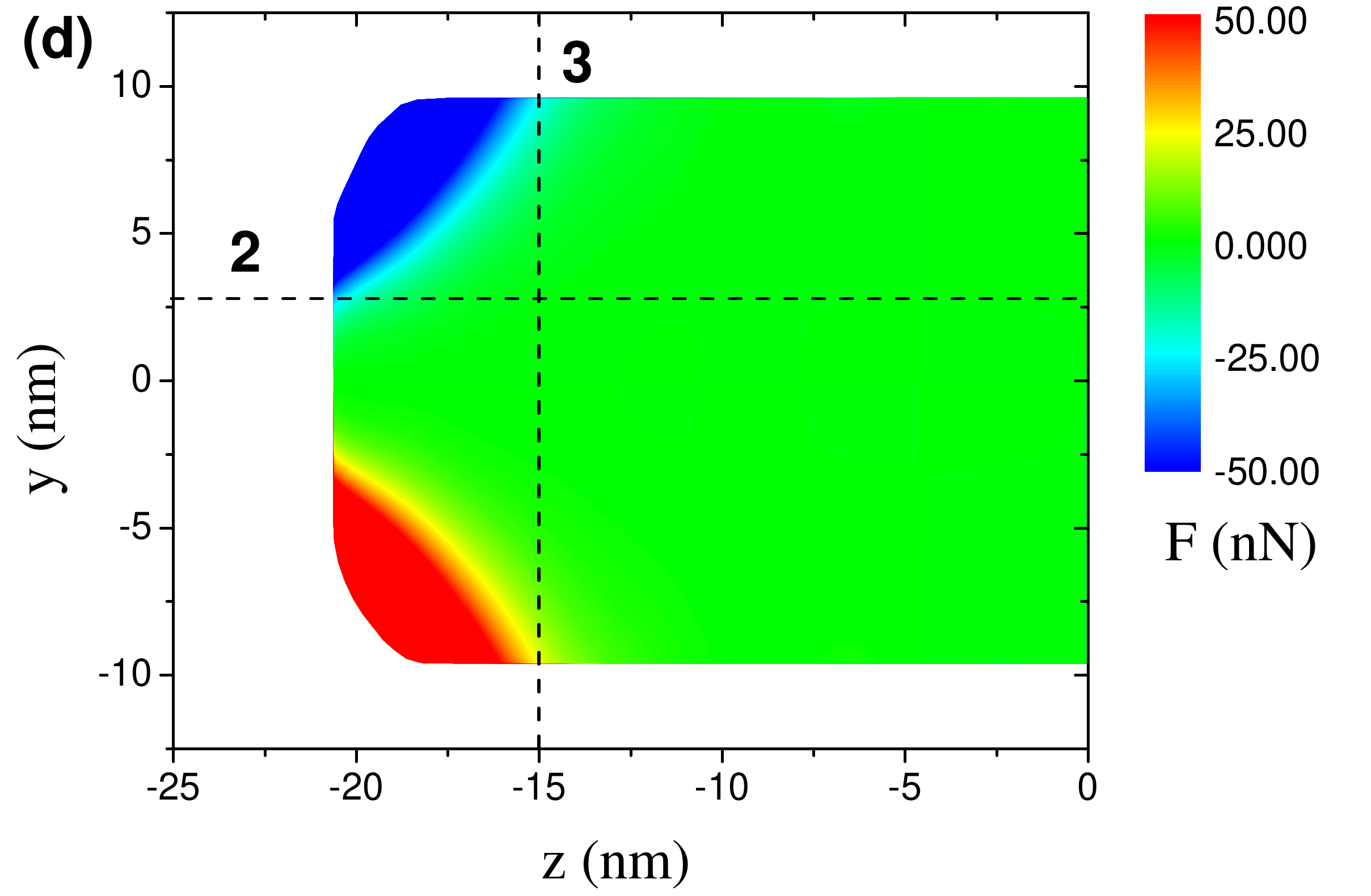}\\
 \caption{\label{fig:2D_f_cons}(Color online) Components $F_z$ and $F_y$ of two-dimensional conservative force in the region of the free tip motion (a,b) and reconstructed components in the region of the engaged tip motion (c,d). Cross-sections 1--3 are illustrated in fig.~\ref{fig:2D_f_cons_cuts}(a)--(c) to highlight agreement between the actual force used in simulation and the reconstructed force.}
\end{figure*}

Simulation parameters for the system (\ref{case2_worksys}) are: $\omega_z = 2\pi\,300$ kHz, $k_z = 40$ N/m, $Q_z = 400$, $\omega_y = 6.3\omega_y$, $k_y = 50k_y$, $Q_2 = 3Q_y$. The driving forces $\mathrm{f}_{z,y}$ are chosen to have the same phase and give maximum free response (when $F_{z,y}\equiv 0$) $A_z=25$ nm and $A_y=12.5$ nm; all four drive frequencies $\omega_{z,y}\pm\delta\omega/2$ are integer multipliers of base frequency $\delta\omega=2\pi\,0.2$ kHz. The engaged height $h$ above the surface is $17$ nm. The model for the component of the tip-surface force perpendicular to the surface is the same vdW-DMT force (\ref{case1_vdW-DMT}) used in the previous case, without the dissipation term $F_z^{dis}$. The model of the force component parallel to the surface is a nonlinear conservative restoring force
\begin{equation}
	F_y(z, y) = - \left(c_1y + c_3y^3\right)e^{-z/\lambda_z}
\label{case2_fmodel_z2}
\end{equation}
where $\lambda_z = 1.5$ nm, $c_1 = 0.22$ N/m and $c_3 = 0.1$ N/m$^3$ are constants. These two components of $\mathbf{F}_{ts}(\mathbf{r})$ are illustrated in fig.~\ref{fig:2D_f_cons}(a)--(b).

\begin{figure*}
\centering
\includegraphics[width=55mm]{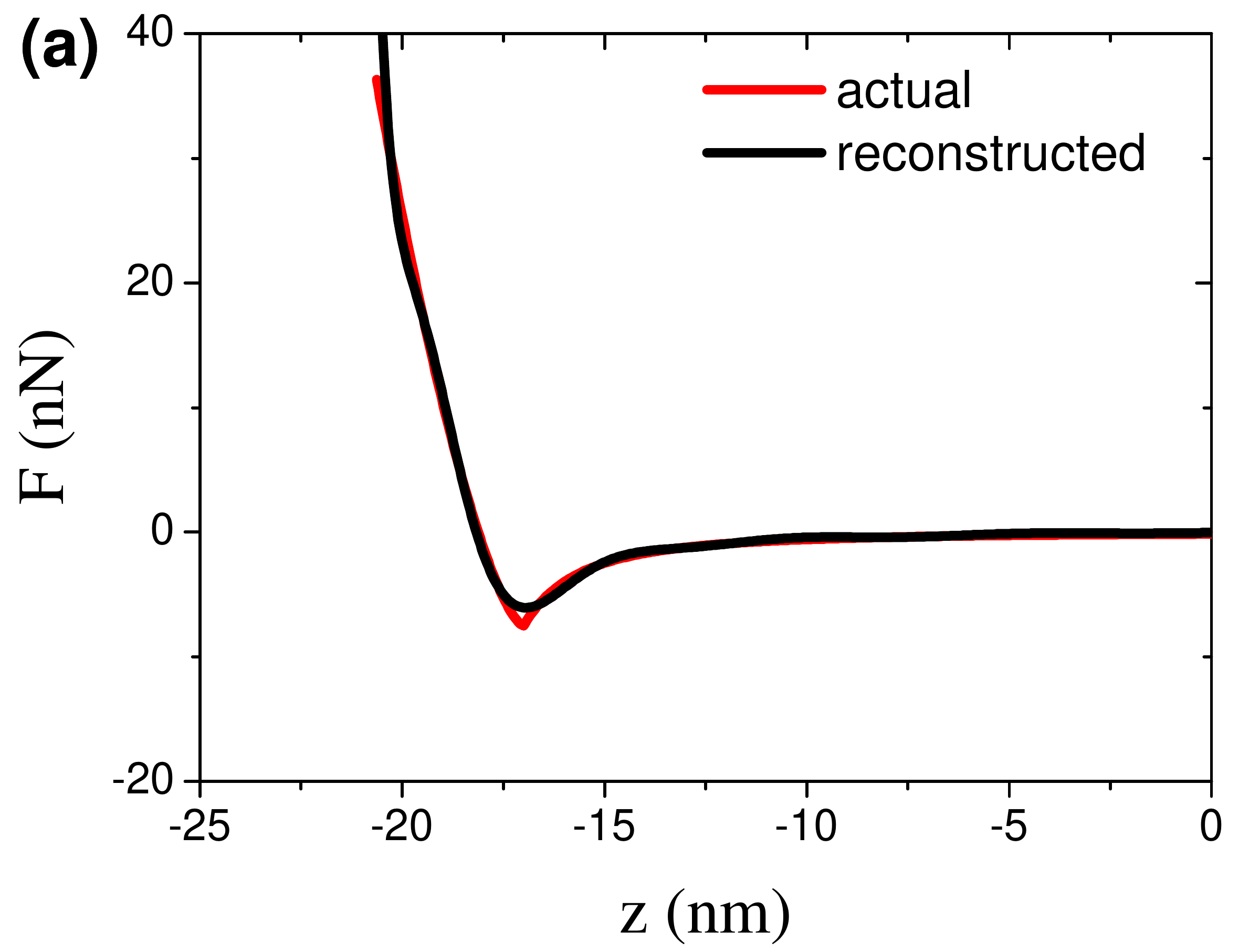}
\includegraphics[width=55mm]{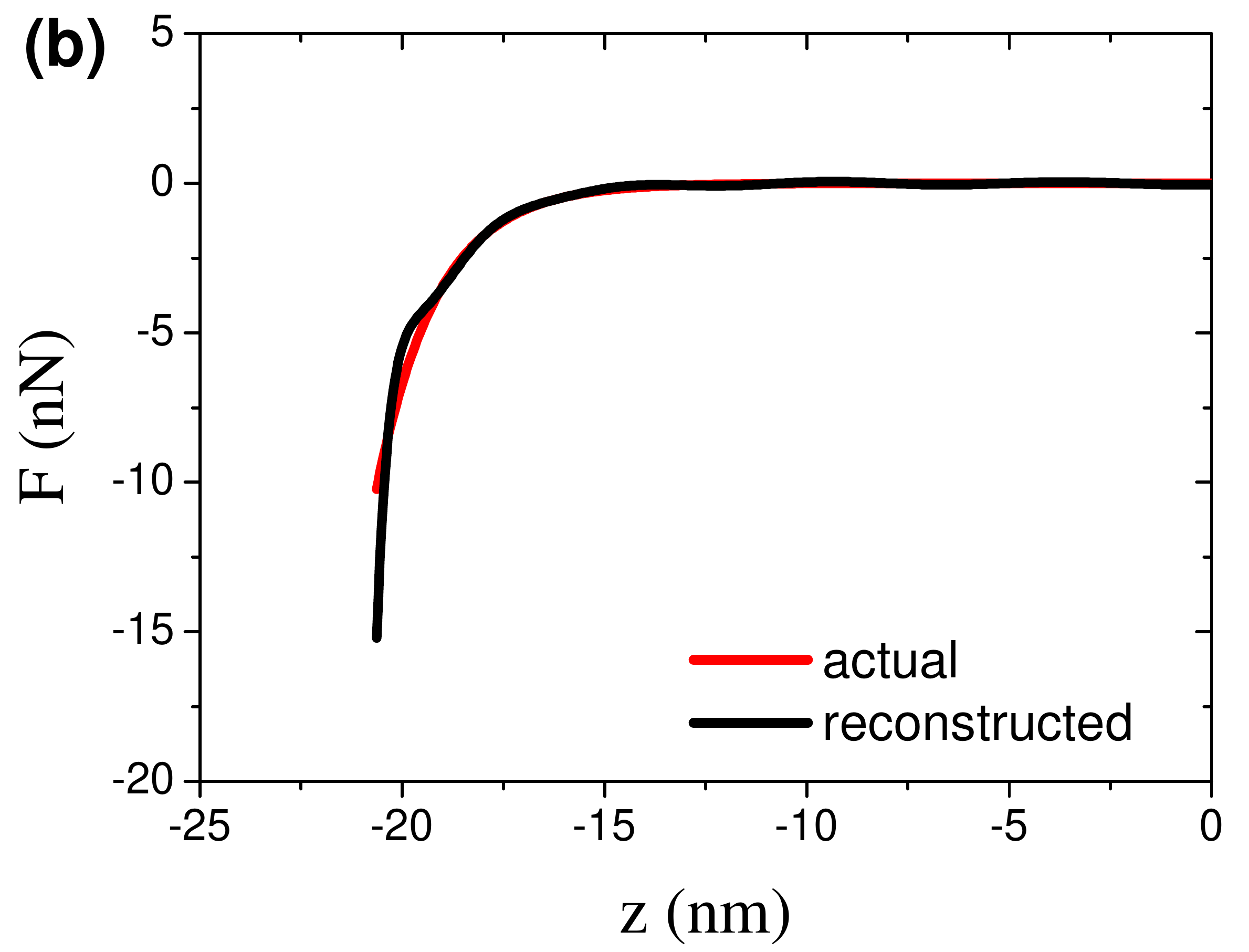}
\includegraphics[width=55mm]{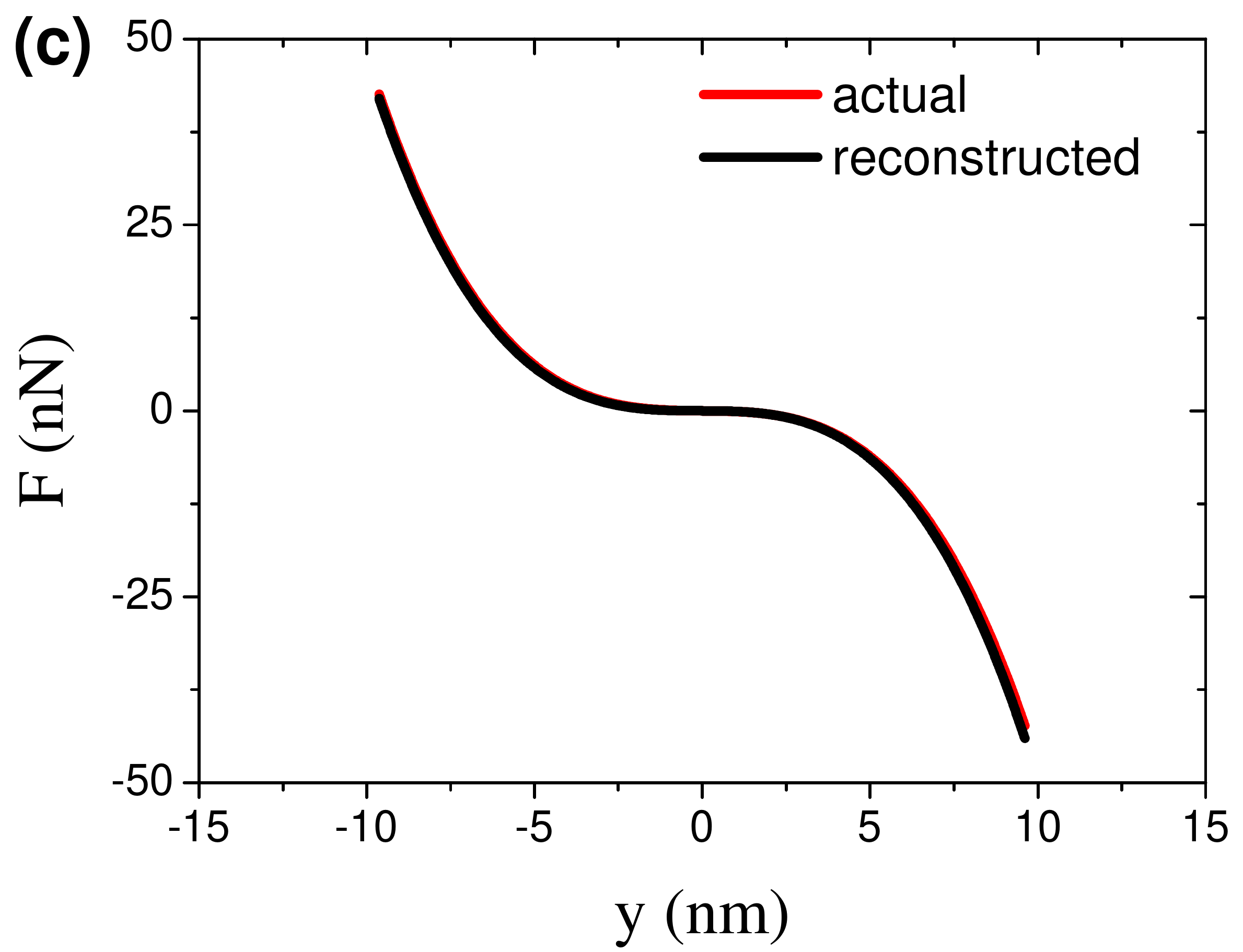}\\
 \caption{\label{fig:2D_f_cons_cuts}(Color online) Cross-sections of reconstructed tip-surface force (black) with comparison to the actual force used in simulation (red) for the cantilever driven at two orthogonal modes. (a) $\tilde F_z(z)$; (b) $\tilde F_y(z,y = 0.2\dot z_{\max})$; (c) $\tilde F_y(z=0.75z_{\min},y)$. Reconstructed force is in a good agreement with underlying model. Perfect agreement is reachable if we assume model for $F_z$ independent of $y$.}
\end{figure*}

Using only $24$ intermodulation peaks in each band for $\hat z$ and $\hat y$, the spectral fitting method reconstructs the two-dimensional vector force field $\mathbf{F}_{ts}$ (\ref{case1_vdW-DMT}) and (\ref{case2_fmodel_z2}) up to the $21^{st}$ power in $z$ and third power in $y$ (fig.~\ref{fig:2D_f_cons}(c)--(d) and \ref{fig:2D_f_cons_cuts}).
The reconstructed force is in good agreement with the actual model and perfect agreement is reachable if we assume the model for $\tilde F_z$ as eq. (\ref{case2_models}) only, independent of $y$.

%-------------------------------------------------------------------------------
\section{\label{sec:conclusions}Concluding remarks}

In this paper we have discussed the basic problem of multimodal ImAFM and we proposed a theoretical framework for reconstructing multidimensional forces using this technique. We demonstrated the possibility of reconstructing tip-surface interactions for two characteristic bimodal cases. We have studied tip-surface interactions that show the most nonlinear behavior in the first degree of freedom, and are linear or cubic in the second degree of freedom. As the numerical results have shown, it is possible to accurately reconstruct dependencies up to a cubic order in the second degree of freedom for the two-dimensional force model using information from only one frequency band. We found that excitation of two flexural modes with two well separated resonances does not allow for a precise reconstruction of a nonlinear damping force using only the information contained in corresponding narrow bands near the resonances. However, the reconstruction nicely captures the overall linear trend, or that of a viscous damping. The use of a second, higher frequency eigenmode allows for reconstruction on a wider region of the phase space of a tip motion, enabling exploration of dissipative interactions inaccessible to the first mode alone for a given maximum amplitude of motion. Additionally, the first eigenmode is found to be more sensitive to dissipation forces acting on the tip. Finally, using simultaneous excitation of two orthogonal modes we can reconstruct nonlinear position-dependent lateral forces simultaneously with vertical forces.  This approach represents a path toward the determination of a vectorial force field by frequency domain multiplexing of the multimodal response of an AFM cantilever.

%-------------------------------------------------------------------------------
\appendix*
%-------------------------------------------------------------------------------
\section{\label{app:coord}Generalized eigencoordinates}

We start from the governing equation for a two-dimensional cantilever
\begin{equation}
	\left(\mathcal{G}_{xy}+\mathcal{G}_t\right)\left[w\left(x,y,t\right)\right]=F(x,y,t)
	\label{supereq}
\end{equation} 
and try to find solution $w\left(x,y,t\right)$ separated in time and space, and expanded into the set of normal modes
\begin{equation}
	w(x,y,t) = \sum\limits_{i=0}^{\infty}\phi_{i}(x,y)q_{i}(t)
	\label{sol}
\end{equation}
Functions $\phi_{i}$ form orthonormal set on the geometrical shape of cantilever $\Omega_c$
\begin{equation}
	\int\limits_{\Omega_c}\phi_{i}\phi_{j}\, \mathrm{d}\Omega_{xy} = \delta_{i}^j
	\label{ortho}
\end{equation}
where $\delta_{i}^j$ is Kronecker delta equals $1$ for $i=j$ and $0$ otherwise.

Inserting solution (\ref{sol}) into eq. (\ref{supereq}) with following multiplication by $\phi_{i}$ and integrating over the plane $\Omega_c$ yields a system of differential equations for the generalized coordinates $q_{i}(t)$
\begin{equation}
	q_{i}\int\limits_{\Omega_c}\phi_{i}\mathcal{G}_{xy}\phi_{i}\, \mathrm{d}\Omega_{xy} + \mathcal{G}_{t}[q_{i}] \int\limits_{\Omega_c}\phi_{i}^{2}\, \mathrm{d}\Omega_{xy} = \int\limits_{\Omega_c}F(\Omega,t)\phi_{ij}\, \mathrm{d}\Omega_{xy}
	\label{supersys}
\end{equation}
here the orthonormal condition (\ref{ortho}) is used. 

Denoting $\mathcal{G}_{i}$ as a differential operator governing motion of $i^{\mathrm{th}}$ generalized coordinate
\begin{equation}
	\mathcal{G}_{i} = k_{i} + m_{i}\mathcal{G}_t
	\label{sysij}
\end{equation} 
where
\begin{equation}
\begin{array}{lcl}
	k_{i} \equiv \int\limits_{\Omega_c}\phi_{i}\mathcal{G}_{xy}\phi_{i}\, \mathrm{d}\Omega_{xy} \\
	m_{i} \equiv \int\limits_{\Omega_c}\phi_{i}^{2}\, \mathrm{d}\Omega_{xy} 
\end{array}
	\label{gij}
\end{equation} 
are effective stiffness and mass of corresponding degree of freedom, and considering damping and inertia
\begin{equation}
	\mathcal{G}_t:=\frac{\partial^2}{\partial t^2} + \gamma\frac{\partial}{\partial t}
	\label{timeoper}
\end{equation}
with some constant $\gamma$ (homogeneous viscous medium damping), we arrive at the final system 
\begin{equation}
	k_i\left(\frac{1}{\omega_i^2}\ddot q_i + \frac{1}{Q_i\omega_i}\dot q_i + q_i\right) = F_{i}\left(t\right) + \mathrm{f}_i\left(t\right)
	\label{supersys2}
\end{equation}
where resonant frequencies $\omega_{j}=\sqrt{k_{i}/m_{i}}$ and quality factors $Q_{i}=\omega_{i}/\gamma$ are introduced. Here the time-dependent forces
\begin{equation}
	F_{i}\left(t\right) + \mathrm{f}_i\left(t\right) := \int\limits_{\Omega_c}\phi_{i}(x,y)F(x,y,t)\, \mathrm{d}\Omega_{xy}
	\label{fij}
\end{equation}
represent anharmonic contribution of the tip-surface interaction and harmonic contribution of the drive, respectively.

%-------------------------------------------------------------------------------
\begin{acknowledgments}
This work is supported by Nordita, DOE, VR VCB 621-2012-2983, the Knut and Allice Wallenberg Foundation, and the Olle Enqvist Foundation. ASdW's work is financially supported by an Unga Forskare grant from the Swedish Research Council.
\end{acknowledgments}

\bibliography{multimode}

\providecommand{\noopsort}[1]{}\providecommand{\singleletter}[1]{#1}%
\begin{thebibliography}{70}%
\makeatletter
\providecommand \@ifxundefined [1]{%
 \@ifx{#1\undefined}
}%
\providecommand \@ifnum [1]{%
 \ifnum #1\expandafter \@firstoftwo
 \else \expandafter \@secondoftwo
 \fi
}%
\providecommand \@ifx [1]{%
 \ifx #1\expandafter \@firstoftwo
 \else \expandafter \@secondoftwo
 \fi
}%
\providecommand \natexlab [1]{#1}%
\providecommand \enquote  [1]{``#1''}%
\providecommand \bibnamefont  [1]{#1}%
\providecommand \bibfnamefont [1]{#1}%
\providecommand \citenamefont [1]{#1}%
\providecommand \href@noop [0]{\@secondoftwo}%
\providecommand \href [0]{\begingroup \@sanitize@url \@href}%
\providecommand \@href[1]{\@@startlink{#1}\@@href}%
\providecommand \@@href[1]{\endgroup#1\@@endlink}%
\providecommand \@sanitize@url [0]{\catcode `\\12\catcode `\$12\catcode
  `\&12\catcode `\#12\catcode `\^12\catcode `\_12\catcode `\%12\relax}%
\providecommand \@@startlink[1]{}%
\providecommand \@@endlink[0]{}%
\providecommand \url  [0]{\begingroup\@sanitize@url \@url }%
\providecommand \@url [1]{\endgroup\@href {#1}{\urlprefix }}%
\providecommand \urlprefix  [0]{URL }%
\providecommand \Eprint [0]{\href }%
\providecommand \doibase [0]{http://dx.doi.org/}%
\providecommand \selectlanguage [0]{\@gobble}%
\providecommand \bibinfo  [0]{\@secondoftwo}%
\providecommand \bibfield  [0]{\@secondoftwo}%
\providecommand \translation [1]{[#1]}%
\providecommand \BibitemOpen [0]{}%
\providecommand \bibitemStop [0]{}%
\providecommand \bibitemNoStop [0]{.\EOS\space}%
\providecommand \EOS [0]{\spacefactor3000\relax}%
\providecommand \BibitemShut  [1]{\csname bibitem#1\endcsname}%
\let\auto@bib@innerbib\@empty
%</preamble>
\bibitem [{\citenamefont {Binnig}\ \emph {et~al.}(1986)\citenamefont {Binnig},
  \citenamefont {Quate},\ and\ \citenamefont {Gerber}}]{AFM1}%
  \BibitemOpen
  \bibfield  {author} {\bibinfo {author} {\bibfnamefont {G.}~\bibnamefont
  {Binnig}}, \bibinfo {author} {\bibfnamefont {C.~F.}\ \bibnamefont {Quate}}, \
  and\ \bibinfo {author} {\bibfnamefont {C.}~\bibnamefont {Gerber}},\ }\href
  {\doibase 10.1103/PhysRevLett.56.930} {\bibfield  {journal} {\bibinfo
  {journal} {Phys. Rev. Lett.}\ }\textbf {\bibinfo {volume} {56}},\ \bibinfo
  {pages} {930} (\bibinfo {year} {1986})}\BibitemShut {NoStop}%
\bibitem [{\citenamefont {Burnham}\ \emph {et~al.}(1993)\citenamefont
  {Burnham}, \citenamefont {Colton},\ and\ \citenamefont
  {Pollock}}]{forceIterpret1}%
  \BibitemOpen
  \bibfield  {author} {\bibinfo {author} {\bibfnamefont {N.~A.}\ \bibnamefont
  {Burnham}}, \bibinfo {author} {\bibfnamefont {R.~J.}\ \bibnamefont {Colton}},
  \ and\ \bibinfo {author} {\bibfnamefont {H.~M.}\ \bibnamefont {Pollock}},\
  }\href {http://stacks.iop.org/0957-4484/4/i=2/a=002} {\bibfield  {journal}
  {\bibinfo  {journal} {Nanotechnology}\ }\textbf {\bibinfo {volume} {4}},\
  \bibinfo {pages} {64} (\bibinfo {year} {1993})}\BibitemShut {NoStop}%
\bibitem [{\citenamefont {Butt}\ \emph {et~al.}(2005)\citenamefont {Butt},
  \citenamefont {Cappella},\ and\ \citenamefont {Kappl}}]{forceIterpret2}%
  \BibitemOpen
  \bibfield  {author} {\bibinfo {author} {\bibfnamefont {H.-J.}\ \bibnamefont
  {Butt}}, \bibinfo {author} {\bibfnamefont {B.}~\bibnamefont {Cappella}}, \
  and\ \bibinfo {author} {\bibfnamefont {M.}~\bibnamefont {Kappl}},\ }\href
  {\doibase 10.1016/j.surfrep.2005.08.003} {\bibfield  {journal} {\bibinfo
  {journal} {Surface Science Reports}\ }\textbf {\bibinfo {volume} {59}},\
  \bibinfo {pages} {1 } (\bibinfo {year} {2005})}\BibitemShut {NoStop}%
\bibitem [{\citenamefont {Hoffmann}\ \emph {et~al.}(2001)\citenamefont
  {Hoffmann}, \citenamefont {Jeffery}, \citenamefont {Pethica}, \citenamefont
  {{\"O}zg{\"u}r~{\"O}zer},\ and\ \citenamefont {Oral}}]{nanoDiss1}%
  \BibitemOpen
  \bibfield  {author} {\bibinfo {author} {\bibfnamefont {P.~M.}\ \bibnamefont
  {Hoffmann}}, \bibinfo {author} {\bibfnamefont {S.}~\bibnamefont {Jeffery}},
  \bibinfo {author} {\bibfnamefont {J.~B.}\ \bibnamefont {Pethica}}, \bibinfo
  {author} {\bibfnamefont {H.~{\"O}.}\ \bibnamefont {{\"O}zg{\"u}r~{\"O}zer}},
  \ and\ \bibinfo {author} {\bibfnamefont {A.}~\bibnamefont {Oral}},\
  }\href@noop {} {\bibfield  {journal} {\bibinfo  {journal} {Phys. Rev. Lett.}\
  }\textbf {\bibinfo {volume} {87}},\ \bibinfo {pages} {265502} (\bibinfo
  {year} {2001})}\BibitemShut {NoStop}%
\bibitem [{\citenamefont {Garc\'ia}\ \emph
  {et~al.}(2006{\natexlab{a}})\citenamefont {Garc\'ia}, \citenamefont
  {G\'omez}, \citenamefont {Martinez}, \citenamefont {Patil}, \citenamefont
  {Dietz},\ and\ \citenamefont {Magerle}}]{nanoDiss2}%
  \BibitemOpen
  \bibfield  {author} {\bibinfo {author} {\bibfnamefont {R.}~\bibnamefont
  {Garc\'ia}}, \bibinfo {author} {\bibfnamefont {C.~J.}\ \bibnamefont
  {G\'omez}}, \bibinfo {author} {\bibfnamefont {N.~F.}\ \bibnamefont
  {Martinez}}, \bibinfo {author} {\bibfnamefont {S.}~\bibnamefont {Patil}},
  \bibinfo {author} {\bibfnamefont {C.}~\bibnamefont {Dietz}}, \ and\ \bibinfo
  {author} {\bibfnamefont {R.}~\bibnamefont {Magerle}},\ }\href@noop {}
  {\bibfield  {journal} {\bibinfo  {journal} {Phys. Rev. Lett.}\ }\textbf
  {\bibinfo {volume} {97}},\ \bibinfo {pages} {016103} (\bibinfo {year}
  {2006}{\natexlab{a}})}\BibitemShut {NoStop}%
\bibitem [{\citenamefont {Negri}\ \emph {et~al.}(2010)\citenamefont {Negri},
  \citenamefont {Manini}, \citenamefont {Vanossi}, \citenamefont {Santoro},\
  and\ \citenamefont {Tosatti}}]{nanoDiss3}%
  \BibitemOpen
  \bibfield  {author} {\bibinfo {author} {\bibfnamefont {C.}~\bibnamefont
  {Negri}}, \bibinfo {author} {\bibfnamefont {N.}~\bibnamefont {Manini}},
  \bibinfo {author} {\bibfnamefont {A.}~\bibnamefont {Vanossi}}, \bibinfo
  {author} {\bibfnamefont {G.~E.}\ \bibnamefont {Santoro}}, \ and\ \bibinfo
  {author} {\bibfnamefont {E.}~\bibnamefont {Tosatti}},\ }\href {\doibase
  10.1103/PhysRevB.81.045417} {\bibfield  {journal} {\bibinfo  {journal} {Phys.
  Rev. B}\ }\textbf {\bibinfo {volume} {81}},\ \bibinfo {pages} {045417}
  (\bibinfo {year} {2010})}\BibitemShut {NoStop}%
\bibitem [{\citenamefont {Saitoh}\ \emph {et~al.}(2010)\citenamefont {Saitoh},
  \citenamefont {Hayashi}, \citenamefont {Shibayama},\ and\ \citenamefont
  {Shirahama}}]{nanoDiss4}%
  \BibitemOpen
  \bibfield  {author} {\bibinfo {author} {\bibfnamefont {K.}~\bibnamefont
  {Saitoh}}, \bibinfo {author} {\bibfnamefont {K.}~\bibnamefont {Hayashi}},
  \bibinfo {author} {\bibfnamefont {Y.}~\bibnamefont {Shibayama}}, \ and\
  \bibinfo {author} {\bibfnamefont {K.}~\bibnamefont {Shirahama}},\ }\href
  {\doibase 10.1103/PhysRevLett.105.236103} {\bibfield  {journal} {\bibinfo
  {journal} {Phys. Rev. Lett.}\ }\textbf {\bibinfo {volume} {105}},\ \bibinfo
  {pages} {236103} (\bibinfo {year} {2010})}\BibitemShut {NoStop}%
\bibitem [{\citenamefont {She}\ and\ \citenamefont
  {Balatsky}(2012)}]{nanoDiss5}%
  \BibitemOpen
  \bibfield  {author} {\bibinfo {author} {\bibfnamefont {J.-H.}\ \bibnamefont
  {She}}\ and\ \bibinfo {author} {\bibfnamefont {A.~V.}\ \bibnamefont
  {Balatsky}},\ }\href {\doibase 10.1103/PhysRevLett.108.136101} {\bibfield
  {journal} {\bibinfo  {journal} {Phys. Rev. Lett.}\ }\textbf {\bibinfo
  {volume} {108}},\ \bibinfo {pages} {136101} (\bibinfo {year}
  {2012})}\BibitemShut {NoStop}%
\bibitem [{\citenamefont {Krotil}\ \emph {et~al.}(1999)\citenamefont {Krotil},
  \citenamefont {Stifter}, \citenamefont {Waschipky}, \citenamefont
  {Weishaupt}, \citenamefont {Hild},\ and\ \citenamefont {Marti}}]{dynAFM1}%
  \BibitemOpen
  \bibfield  {author} {\bibinfo {author} {\bibfnamefont {H.~U.}\ \bibnamefont
  {Krotil}}, \bibinfo {author} {\bibfnamefont {T.}~\bibnamefont {Stifter}},
  \bibinfo {author} {\bibfnamefont {H.}~\bibnamefont {Waschipky}}, \bibinfo
  {author} {\bibfnamefont {K.}~\bibnamefont {Weishaupt}}, \bibinfo {author}
  {\bibfnamefont {S.}~\bibnamefont {Hild}}, \ and\ \bibinfo {author}
  {\bibfnamefont {O.}~\bibnamefont {Marti}},\ }\href@noop {} {\bibfield
  {journal} {\bibinfo  {journal} {Surf. Interface Anal.}\ }\textbf {\bibinfo
  {volume} {27}},\ \bibinfo {pages} {336–} (\bibinfo {year}
  {1999})}\BibitemShut {NoStop}%
\bibitem [{\citenamefont {Garc\'ia}\ and\ \citenamefont
  {Perez}(2002)}]{dynAFM2}%
  \BibitemOpen
  \bibfield  {author} {\bibinfo {author} {\bibfnamefont {R.}~\bibnamefont
  {Garc\'ia}}\ and\ \bibinfo {author} {\bibfnamefont {R.}~\bibnamefont
  {Perez}},\ }\href@noop {} {\bibfield  {journal} {\bibinfo  {journal} {Surf.
  Sci. Rep.}\ }\textbf {\bibinfo {volume} {47}},\ \bibinfo {pages} {197}
  (\bibinfo {year} {2002})}\BibitemShut {NoStop}%
\bibitem [{\citenamefont {Giessibl}(2003)}]{dynAFM3}%
  \BibitemOpen
  \bibfield  {author} {\bibinfo {author} {\bibfnamefont {F.~J.}\ \bibnamefont
  {Giessibl}},\ }\href@noop {} {\bibfield  {journal} {\bibinfo  {journal} {Rev.
  Mod. Phys.}\ }\textbf {\bibinfo {volume} {75}},\ \bibinfo {pages} {949}
  (\bibinfo {year} {2003})}\BibitemShut {NoStop}%
\bibitem [{\citenamefont {Sader}\ \emph {et~al.}(2005)\citenamefont {Sader},
  \citenamefont {Uchihashi}, \citenamefont {Higgins}, \citenamefont {Farrell},
  \citenamefont {Nakayama},\ and\ \citenamefont {Jarvis}}]{dynAFM4}%
  \BibitemOpen
  \bibfield  {author} {\bibinfo {author} {\bibfnamefont {J.~E.}\ \bibnamefont
  {Sader}}, \bibinfo {author} {\bibfnamefont {T.}~\bibnamefont {Uchihashi}},
  \bibinfo {author} {\bibfnamefont {M.~J.}\ \bibnamefont {Higgins}}, \bibinfo
  {author} {\bibfnamefont {A.}~\bibnamefont {Farrell}}, \bibinfo {author}
  {\bibfnamefont {Y.}~\bibnamefont {Nakayama}}, \ and\ \bibinfo {author}
  {\bibfnamefont {S.~P.}\ \bibnamefont {Jarvis}},\ }\href@noop {} {\bibfield
  {journal} {\bibinfo  {journal} {Nanotechnology}\ }\textbf {\bibinfo {volume}
  {16}},\ \bibinfo {pages} {94} (\bibinfo {year} {2005})}\BibitemShut {NoStop}%
\bibitem [{\citenamefont {Albers}\ \emph {et~al.}(2009)\citenamefont {Albers},
  \citenamefont {Schwendemann}, \citenamefont {Baykara}, \citenamefont {Pilet},
  \citenamefont {Liebmann}, \citenamefont {Altman},\ and\ \citenamefont
  {Schwarz}}]{dynAFM5}%
  \BibitemOpen
  \bibfield  {author} {\bibinfo {author} {\bibfnamefont {B.~J.}\ \bibnamefont
  {Albers}}, \bibinfo {author} {\bibfnamefont {T.~C.}\ \bibnamefont
  {Schwendemann}}, \bibinfo {author} {\bibfnamefont {M.~Z.}\ \bibnamefont
  {Baykara}}, \bibinfo {author} {\bibfnamefont {N.}~\bibnamefont {Pilet}},
  \bibinfo {author} {\bibfnamefont {M.}~\bibnamefont {Liebmann}}, \bibinfo
  {author} {\bibfnamefont {E.~I.}\ \bibnamefont {Altman}}, \ and\ \bibinfo
  {author} {\bibfnamefont {U.~D.}\ \bibnamefont {Schwarz}},\ }\href@noop {}
  {\bibfield  {journal} {\bibinfo  {journal} {Nanotechnology}\ }\textbf
  {\bibinfo {volume} {20}},\ \bibinfo {pages} {264002} (\bibinfo {year}
  {2009})}\BibitemShut {NoStop}%
\bibitem [{\citenamefont {Gross}\ \emph {et~al.}(2009)\citenamefont {Gross},
  \citenamefont {Mohn}, \citenamefont {Moll}, \citenamefont {Liljeroth},\ and\
  \citenamefont {Meyer}}]{dynAFM6}%
  \BibitemOpen
  \bibfield  {author} {\bibinfo {author} {\bibfnamefont {L.}~\bibnamefont
  {Gross}}, \bibinfo {author} {\bibfnamefont {F.}~\bibnamefont {Mohn}},
  \bibinfo {author} {\bibfnamefont {N.}~\bibnamefont {Moll}}, \bibinfo {author}
  {\bibfnamefont {P.}~\bibnamefont {Liljeroth}}, \ and\ \bibinfo {author}
  {\bibfnamefont {G.}~\bibnamefont {Meyer}},\ }\href@noop {} {\bibfield
  {journal} {\bibinfo  {journal} {Science}\ }\textbf {\bibinfo {volume}
  {325}},\ \bibinfo {pages} {1110} (\bibinfo {year} {2009})}\BibitemShut
  {NoStop}%
\bibitem [{\citenamefont {Solares}\ and\ \citenamefont
  {Chawla}(2010)}]{dynAFM7}%
  \BibitemOpen
  \bibfield  {author} {\bibinfo {author} {\bibfnamefont {S.~D.}\ \bibnamefont
  {Solares}}\ and\ \bibinfo {author} {\bibfnamefont {G.}~\bibnamefont
  {Chawla}},\ }\href@noop {} {\bibfield  {journal} {\bibinfo  {journal} {J.
  Appl. Phys.}\ }\textbf {\bibinfo {volume} {108}},\ \bibinfo {pages} {054901}
  (\bibinfo {year} {2010})}\BibitemShut {NoStop}%
\bibitem [{\citenamefont {Albrecht}\ \emph {et~al.}(1991)\citenamefont
  {Albrecht}, \citenamefont {Grutter}, \citenamefont {Horne},\ and\
  \citenamefont {Rugar}}]{like_imafm0}%
  \BibitemOpen
  \bibfield  {author} {\bibinfo {author} {\bibfnamefont {T.~R.}\ \bibnamefont
  {Albrecht}}, \bibinfo {author} {\bibfnamefont {P.}~\bibnamefont {Grutter}},
  \bibinfo {author} {\bibfnamefont {D.}~\bibnamefont {Horne}}, \ and\ \bibinfo
  {author} {\bibfnamefont {D.}~\bibnamefont {Rugar}},\ }\href@noop {}
  {\bibfield  {journal} {\bibinfo  {journal} {J. Appl. Phys.}\ }\textbf
  {\bibinfo {volume} {69}},\ \bibinfo {pages} {668} (\bibinfo {year}
  {1991})}\BibitemShut {NoStop}%
\bibitem [{\citenamefont {Stark}\ \emph {et~al.}(2002)\citenamefont {Stark},
  \citenamefont {Stark}, \citenamefont {Heckl},\ and\ \citenamefont
  {Guckenberger}}]{like_imafm1}%
  \BibitemOpen
  \bibfield  {author} {\bibinfo {author} {\bibfnamefont {M.}~\bibnamefont
  {Stark}}, \bibinfo {author} {\bibfnamefont {R.~W.}\ \bibnamefont {Stark}},
  \bibinfo {author} {\bibfnamefont {W.~M.}\ \bibnamefont {Heckl}}, \ and\
  \bibinfo {author} {\bibfnamefont {R.}~\bibnamefont {Guckenberger}},\
  }\href@noop {} {\bibfield  {journal} {\bibinfo  {journal} {Proc. Natl. Acad.
  Sci. USA}\ }\textbf {\bibinfo {volume} {99}},\ \bibinfo {pages} {8473}
  (\bibinfo {year} {2002})}\BibitemShut {NoStop}%
\bibitem [{\citenamefont {Legleiter}\ \emph {et~al.}(2006)\citenamefont
  {Legleiter}, \citenamefont {Park}, \citenamefont {Cusick},\ and\
  \citenamefont {Kowalewski}}]{like_imafm2}%
  \BibitemOpen
  \bibfield  {author} {\bibinfo {author} {\bibfnamefont {J.}~\bibnamefont
  {Legleiter}}, \bibinfo {author} {\bibfnamefont {M.}~\bibnamefont {Park}},
  \bibinfo {author} {\bibfnamefont {B.}~\bibnamefont {Cusick}}, \ and\ \bibinfo
  {author} {\bibfnamefont {T.}~\bibnamefont {Kowalewski}},\ }\href@noop {}
  {\bibfield  {journal} {\bibinfo  {journal} {Proc. Natl. Acad. Sci. USA}\
  }\textbf {\bibinfo {volume} {103}},\ \bibinfo {pages} {4813} (\bibinfo {year}
  {2006})}\BibitemShut {NoStop}%
\bibitem [{\citenamefont {Lee}\ and\ \citenamefont {Jhe}(2006)}]{like_imafm3}%
  \BibitemOpen
  \bibfield  {author} {\bibinfo {author} {\bibfnamefont {M.}~\bibnamefont
  {Lee}}\ and\ \bibinfo {author} {\bibfnamefont {W.}~\bibnamefont {Jhe}},\
  }\href@noop {} {\bibfield  {journal} {\bibinfo  {journal} {Phys. Rev. Lett.}\
  }\textbf {\bibinfo {volume} {97}},\ \bibinfo {pages} {036104} (\bibinfo
  {year} {2006})}\BibitemShut {NoStop}%
\bibitem [{\citenamefont {Jesse}\ \emph {et~al.}(2007)\citenamefont {Jesse},
  \citenamefont {Kalinin}, \citenamefont {Proksch}, \citenamefont {Baddorf},\
  and\ \citenamefont {Rodriguez}}]{like_imafm4}%
  \BibitemOpen
  \bibfield  {author} {\bibinfo {author} {\bibfnamefont {S.}~\bibnamefont
  {Jesse}}, \bibinfo {author} {\bibfnamefont {S.~V.}\ \bibnamefont {Kalinin}},
  \bibinfo {author} {\bibfnamefont {R.}~\bibnamefont {Proksch}}, \bibinfo
  {author} {\bibfnamefont {A.~P.}\ \bibnamefont {Baddorf}}, \ and\ \bibinfo
  {author} {\bibfnamefont {B.~J.}\ \bibnamefont {Rodriguez}},\ }\href@noop {}
  {\bibfield  {journal} {\bibinfo  {journal} {Nanotechnology}\ }\textbf
  {\bibinfo {volume} {18}},\ \bibinfo {pages} {435503} (\bibinfo {year}
  {2007})}\BibitemShut {NoStop}%
\bibitem [{\citenamefont {Sahin}\ \emph {et~al.}(2007)\citenamefont {Sahin},
  \citenamefont {Magonov}, \citenamefont {Su}, \citenamefont {Quate},\ and\
  \citenamefont {Solgaard}}]{like_imafm5}%
  \BibitemOpen
  \bibfield  {author} {\bibinfo {author} {\bibfnamefont {O.}~\bibnamefont
  {Sahin}}, \bibinfo {author} {\bibfnamefont {S.}~\bibnamefont {Magonov}},
  \bibinfo {author} {\bibfnamefont {C.}~\bibnamefont {Su}}, \bibinfo {author}
  {\bibfnamefont {C.~F.}\ \bibnamefont {Quate}}, \ and\ \bibinfo {author}
  {\bibfnamefont {O.}~\bibnamefont {Solgaard}},\ }\href@noop {} {\bibfield
  {journal} {\bibinfo  {journal} {Nature Nanotechnol.}\ }\textbf {\bibinfo
  {volume} {2}},\ \bibinfo {pages} {507} (\bibinfo {year} {2007})}\BibitemShut
  {NoStop}%
\bibitem [{\citenamefont {Rodriguez}\ \emph {et~al.}(2007)\citenamefont
  {Rodriguez}, \citenamefont {Callahan}, \citenamefont {Kalinin},\ and\
  \citenamefont {Proksch}}]{like_imafm6}%
  \BibitemOpen
  \bibfield  {author} {\bibinfo {author} {\bibfnamefont {B.~J.}\ \bibnamefont
  {Rodriguez}}, \bibinfo {author} {\bibfnamefont {C.}~\bibnamefont {Callahan}},
  \bibinfo {author} {\bibfnamefont {S.~V.}\ \bibnamefont {Kalinin}}, \ and\
  \bibinfo {author} {\bibfnamefont {R.}~\bibnamefont {Proksch}},\ }\href@noop
  {} {\bibfield  {journal} {\bibinfo  {journal} {J. Appl. Phys.}\ }\textbf
  {\bibinfo {volume} {18}},\ \bibinfo {pages} {475504} (\bibinfo {year}
  {2007})}\BibitemShut {NoStop}%
\bibitem [{\citenamefont {Labuda}\ \emph {et~al.}(2011)\citenamefont {Labuda},
  \citenamefont {Miyahara}, \citenamefont {Cockins},\ and\ \citenamefont
  {Gr{\"u}tter}}]{like_imafm7}%
  \BibitemOpen
  \bibfield  {author} {\bibinfo {author} {\bibfnamefont {A.}~\bibnamefont
  {Labuda}}, \bibinfo {author} {\bibfnamefont {Y.}~\bibnamefont {Miyahara}},
  \bibinfo {author} {\bibfnamefont {L.}~\bibnamefont {Cockins}}, \ and\
  \bibinfo {author} {\bibfnamefont {P.~H.}\ \bibnamefont {Gr{\"u}tter}},\
  }\href@noop {} {\bibfield  {journal} {\bibinfo  {journal} {Phys. Rev. B}\
  }\textbf {\bibinfo {volume} {84}},\ \bibinfo {pages} {125433} (\bibinfo
  {year} {2011})}\BibitemShut {NoStop}%
\bibitem [{\citenamefont {Sarioglu}\ \emph {et~al.}(2012)\citenamefont
  {Sarioglu}, \citenamefont {Magonov},\ and\ \citenamefont
  {Solgaard}}]{like_imafm8}%
  \BibitemOpen
  \bibfield  {author} {\bibinfo {author} {\bibfnamefont {A.~F.}\ \bibnamefont
  {Sarioglu}}, \bibinfo {author} {\bibfnamefont {S.}~\bibnamefont {Magonov}}, \
  and\ \bibinfo {author} {\bibfnamefont {O.}~\bibnamefont {Solgaard}},\
  }\href@noop {} {\bibfield  {journal} {\bibinfo  {journal} {Appl. Phys.
  Lett.}\ }\textbf {\bibinfo {volume} {100}},\ \bibinfo {pages} {053109}
  (\bibinfo {year} {2012})}\BibitemShut {NoStop}%
\bibitem [{\citenamefont {Serra-Garc\'ia}\ \emph {et~al.}(2012)\citenamefont
  {Serra-Garc\'ia}, \citenamefont {P\'erez-Murano},\ and\ \citenamefont
  {San~Paulo}}]{like_imafm9}%
  \BibitemOpen
  \bibfield  {author} {\bibinfo {author} {\bibfnamefont {M.}~\bibnamefont
  {Serra-Garc\'ia}}, \bibinfo {author} {\bibfnamefont {F.}~\bibnamefont
  {P\'erez-Murano}}, \ and\ \bibinfo {author} {\bibfnamefont {A.}~\bibnamefont
  {San~Paulo}},\ }\href@noop {} {\bibfield  {journal} {\bibinfo  {journal}
  {Phys. Rev. B}\ }\textbf {\bibinfo {volume} {85}},\ \bibinfo {pages} {035433}
  (\bibinfo {year} {2012})}\BibitemShut {NoStop}%
\bibitem [{\citenamefont {Platz}\ \emph {et~al.}(2008)\citenamefont {Platz},
  \citenamefont {Thol\'en}, \citenamefont {Pesen},\ and\ \citenamefont
  {Haviland}}]{imafm1}%
  \BibitemOpen
  \bibfield  {author} {\bibinfo {author} {\bibfnamefont {D.}~\bibnamefont
  {Platz}}, \bibinfo {author} {\bibfnamefont {E.~A.}\ \bibnamefont {Thol\'en}},
  \bibinfo {author} {\bibfnamefont {D.}~\bibnamefont {Pesen}}, \ and\ \bibinfo
  {author} {\bibfnamefont {D.~B.}\ \bibnamefont {Haviland}},\ }\href@noop {}
  {\bibfield  {journal} {\bibinfo  {journal} {Appl. Phys. Lett.}\ }\textbf
  {\bibinfo {volume} {92}},\ \bibinfo {pages} {153106} (\bibinfo {year}
  {2008})}\BibitemShut {NoStop}%
\bibitem [{\citenamefont {Hutter}\ \emph {et~al.}(2010)\citenamefont {Hutter},
  \citenamefont {Platz}, \citenamefont {Thol\'en}, \citenamefont {Hansson},\
  and\ \citenamefont {Haviland}}]{imafm2}%
  \BibitemOpen
  \bibfield  {author} {\bibinfo {author} {\bibfnamefont {C.}~\bibnamefont
  {Hutter}}, \bibinfo {author} {\bibfnamefont {D.}~\bibnamefont {Platz}},
  \bibinfo {author} {\bibfnamefont {E.~A.}\ \bibnamefont {Thol\'en}}, \bibinfo
  {author} {\bibfnamefont {T.~H.}\ \bibnamefont {Hansson}}, \ and\ \bibinfo
  {author} {\bibfnamefont {D.~B.}\ \bibnamefont {Haviland}},\ }\href@noop {}
  {\bibfield  {journal} {\bibinfo  {journal} {Phys. Rev. Lett.}\ }\textbf
  {\bibinfo {volume} {104}},\ \bibinfo {pages} {050801} (\bibinfo {year}
  {2010})}\BibitemShut {NoStop}%
\bibitem [{\citenamefont {Forchheimer}\ \emph {et~al.}(2012)\citenamefont
  {Forchheimer}, \citenamefont {Platz}, \citenamefont {Thol\'en},\ and\
  \citenamefont {Haviland}}]{imafm3}%
  \BibitemOpen
  \bibfield  {author} {\bibinfo {author} {\bibfnamefont {D.}~\bibnamefont
  {Forchheimer}}, \bibinfo {author} {\bibfnamefont {D.}~\bibnamefont {Platz}},
  \bibinfo {author} {\bibfnamefont {E.~A.}\ \bibnamefont {Thol\'en}}, \ and\
  \bibinfo {author} {\bibfnamefont {D.~B.}\ \bibnamefont {Haviland}},\
  }\href@noop {} {\bibfield  {journal} {\bibinfo  {journal} {Phys. Rev. B}\
  }\textbf {\bibinfo {volume} {85}},\ \bibinfo {pages} {195449} (\bibinfo
  {year} {2012})}\BibitemShut {NoStop}%
\bibitem [{\citenamefont {Helmholtz}(1895)}]{combtones}%
  \BibitemOpen
  \bibfield  {author} {\bibinfo {author} {\bibfnamefont {H.}~\bibnamefont
  {Helmholtz}},\ }\href@noop {} {\emph {\bibinfo {title} {Sensations of
  tone}}}\ (\bibinfo  {publisher} {Logmans Green and Co. New York N. Y.},\
  \bibinfo {year} {1895})\BibitemShut {NoStop}%
\bibitem [{\citenamefont {Stark}\ and\ \citenamefont {Heckl}(2000)}]{multi1}%
  \BibitemOpen
  \bibfield  {author} {\bibinfo {author} {\bibfnamefont {R.~W.}\ \bibnamefont
  {Stark}}\ and\ \bibinfo {author} {\bibfnamefont {W.~M.}\ \bibnamefont
  {Heckl}},\ }\href@noop {} {\bibfield  {journal} {\bibinfo  {journal} {Surf.
  Sci.}\ }\textbf {\bibinfo {volume} {457}},\ \bibinfo {pages} {219} (\bibinfo
  {year} {2000})}\BibitemShut {NoStop}%
\bibitem [{\citenamefont {Proksch}(2006)}]{multi2}%
  \BibitemOpen
  \bibfield  {author} {\bibinfo {author} {\bibfnamefont {R.}~\bibnamefont
  {Proksch}},\ }\href@noop {} {\bibfield  {journal} {\bibinfo  {journal} {Appl.
  Phys. Lett.}\ }\textbf {\bibinfo {volume} {89}},\ \bibinfo {pages} {113121}
  (\bibinfo {year} {2006})}\BibitemShut {NoStop}%
\bibitem [{\citenamefont {Martinez}\ \emph {et~al.}(2006)\citenamefont
  {Martinez}, \citenamefont {Patil}, \citenamefont {Lozano},\ and\
  \citenamefont {Garc\'ia}}]{multi3}%
  \BibitemOpen
  \bibfield  {author} {\bibinfo {author} {\bibfnamefont {N.~F.}\ \bibnamefont
  {Martinez}}, \bibinfo {author} {\bibfnamefont {S.}~\bibnamefont {Patil}},
  \bibinfo {author} {\bibfnamefont {J.~R.}\ \bibnamefont {Lozano}}, \ and\
  \bibinfo {author} {\bibfnamefont {R.}~\bibnamefont {Garc\'ia}},\ }\href@noop
  {} {\bibfield  {journal} {\bibinfo  {journal} {Appl. Phys. Lett.}\ }\textbf
  {\bibinfo {volume} {89}},\ \bibinfo {pages} {153115} (\bibinfo {year}
  {2006})}\BibitemShut {NoStop}%
\bibitem [{\citenamefont {Rupp}\ \emph {et~al.}(2007)\citenamefont {Rupp},
  \citenamefont {Rabe}, \citenamefont {Hirsekorn},\ and\ \citenamefont
  {Arnold}}]{multi4}%
  \BibitemOpen
  \bibfield  {author} {\bibinfo {author} {\bibfnamefont {D.}~\bibnamefont
  {Rupp}}, \bibinfo {author} {\bibfnamefont {U.}~\bibnamefont {Rabe}}, \bibinfo
  {author} {\bibfnamefont {S.}~\bibnamefont {Hirsekorn}}, \ and\ \bibinfo
  {author} {\bibfnamefont {W.}~\bibnamefont {Arnold}},\ }\href@noop {}
  {\bibfield  {journal} {\bibinfo  {journal} {J. Phys. D: Appl. Phys.}\
  }\textbf {\bibinfo {volume} {40}},\ \bibinfo {pages} {7136} (\bibinfo {year}
  {2007})}\BibitemShut {NoStop}%
\bibitem [{\citenamefont {Lozano}\ and\ \citenamefont
  {Garc\'ia}(2008)}]{multi5}%
  \BibitemOpen
  \bibfield  {author} {\bibinfo {author} {\bibfnamefont {J.~R.}\ \bibnamefont
  {Lozano}}\ and\ \bibinfo {author} {\bibfnamefont {R.}~\bibnamefont
  {Garc\'ia}},\ }\href@noop {} {\bibfield  {journal} {\bibinfo  {journal}
  {Phys. Rev. Lett.}\ }\textbf {\bibinfo {volume} {100}},\ \bibinfo {pages}
  {076102} (\bibinfo {year} {2008})}\BibitemShut {NoStop}%
\bibitem [{\citenamefont {Lozano}\ and\ \citenamefont {Garcia}(2009)}]{multi6}%
  \BibitemOpen
  \bibfield  {author} {\bibinfo {author} {\bibfnamefont {J.~R.}\ \bibnamefont
  {Lozano}}\ and\ \bibinfo {author} {\bibfnamefont {R.}~\bibnamefont
  {Garcia}},\ }\href@noop {} {\bibfield  {journal} {\bibinfo  {journal} {Phys.
  Rev. B}\ }\textbf {\bibinfo {volume} {79}},\ \bibinfo {pages} {014110}
  (\bibinfo {year} {2009})}\BibitemShut {NoStop}%
\bibitem [{\citenamefont {Martinez-Martin}\ \emph {et~al.}(2011)\citenamefont
  {Martinez-Martin}, \citenamefont {Herruzo}, \citenamefont {Dietz},
  \citenamefont {Gomez-Herrero},\ and\ \citenamefont {Garc\'ia}}]{multi8}%
  \BibitemOpen
  \bibfield  {author} {\bibinfo {author} {\bibfnamefont {D.}~\bibnamefont
  {Martinez-Martin}}, \bibinfo {author} {\bibfnamefont {E.~T.}\ \bibnamefont
  {Herruzo}}, \bibinfo {author} {\bibfnamefont {C.}~\bibnamefont {Dietz}},
  \bibinfo {author} {\bibfnamefont {J.}~\bibnamefont {Gomez-Herrero}}, \ and\
  \bibinfo {author} {\bibfnamefont {R.}~\bibnamefont {Garc\'ia}},\ }\href@noop
  {} {\bibfield  {journal} {\bibinfo  {journal} {Phys. Rev. Lett.}\ }\textbf
  {\bibinfo {volume} {106}},\ \bibinfo {pages} {198101} (\bibinfo {year}
  {2011})}\BibitemShut {NoStop}%
\bibitem [{\citenamefont {Yurtsever}\ \emph {et~al.}(2008)\citenamefont
  {Yurtsever}, \citenamefont {Gigler},\ and\ \citenamefont {Stark}}]{tors0}%
  \BibitemOpen
  \bibfield  {author} {\bibinfo {author} {\bibfnamefont {A.}~\bibnamefont
  {Yurtsever}}, \bibinfo {author} {\bibfnamefont {A.~M.}\ \bibnamefont
  {Gigler}}, \ and\ \bibinfo {author} {\bibfnamefont {R.~W.}\ \bibnamefont
  {Stark}},\ }\href@noop {} {\bibfield  {journal} {\bibinfo  {journal} {J. of
  Phys.}\ }\textbf {\bibinfo {volume} {100}},\ \bibinfo {pages} {052033}
  (\bibinfo {year} {2008})}\BibitemShut {NoStop}%
\bibitem [{\citenamefont {Hakari}\ \emph {et~al.}(2011)\citenamefont {Hakari},
  \citenamefont {Sekiguchi}, \citenamefont {Osada}, \citenamefont {Kishimoto},
  \citenamefont {Afrin},\ and\ \citenamefont {Ikai}}]{tors1}%
  \BibitemOpen
  \bibfield  {author} {\bibinfo {author} {\bibfnamefont {T.}~\bibnamefont
  {Hakari}}, \bibinfo {author} {\bibfnamefont {H.}~\bibnamefont {Sekiguchi}},
  \bibinfo {author} {\bibfnamefont {T.}~\bibnamefont {Osada}}, \bibinfo
  {author} {\bibfnamefont {K.}~\bibnamefont {Kishimoto}}, \bibinfo {author}
  {\bibfnamefont {R.}~\bibnamefont {Afrin}}, \ and\ \bibinfo {author}
  {\bibfnamefont {A.}~\bibnamefont {Ikai}},\ }\href@noop {} {\bibfield
  {journal} {\bibinfo  {journal} {Cytoskeleton}\ }\textbf {\bibinfo {volume}
  {68}},\ \bibinfo {pages} {628} (\bibinfo {year} {2011})}\BibitemShut
  {NoStop}%
\bibitem [{\citenamefont {Bulgarevich}\ \emph {et~al.}(2007)\citenamefont
  {Bulgarevich}, \citenamefont {Mitsui},\ and\ \citenamefont
  {Arakawa}}]{tors2}%
  \BibitemOpen
  \bibfield  {author} {\bibinfo {author} {\bibfnamefont {D.~S.}\ \bibnamefont
  {Bulgarevich}}, \bibinfo {author} {\bibfnamefont {K.}~\bibnamefont {Mitsui}},
  \ and\ \bibinfo {author} {\bibfnamefont {H.}~\bibnamefont {Arakawa}},\
  }\href@noop {} {\bibfield  {journal} {\bibinfo  {journal} {J. Phys.: Conf.
  Ser.}\ }\textbf {\bibinfo {volume} {61}},\ \bibinfo {pages} {170} (\bibinfo
  {year} {2007})}\BibitemShut {NoStop}%
\bibitem [{\citenamefont {Gigler}\ \emph {et~al.}(2012)\citenamefont {Gigler},
  \citenamefont {Dietz}, \citenamefont {Baumann}, \citenamefont {Martinez},
  \citenamefont {Garc\'ia},\ and\ \citenamefont {Stark}}]{multi10}%
  \BibitemOpen
  \bibfield  {author} {\bibinfo {author} {\bibfnamefont {A.~M.}\ \bibnamefont
  {Gigler}}, \bibinfo {author} {\bibfnamefont {C.}~\bibnamefont {Dietz}},
  \bibinfo {author} {\bibfnamefont {M.}~\bibnamefont {Baumann}}, \bibinfo
  {author} {\bibfnamefont {N.~F.}\ \bibnamefont {Martinez}}, \bibinfo {author}
  {\bibfnamefont {R.}~\bibnamefont {Garc\'ia}}, \ and\ \bibinfo {author}
  {\bibfnamefont {R.~W.}\ \bibnamefont {Stark}},\ }\href@noop {} {\bibfield
  {journal} {\bibinfo  {journal} {Beilstein J Nanotechnol.}\ }\textbf {\bibinfo
  {volume} {3}},\ \bibinfo {pages} {456} (\bibinfo {year} {2012})}\BibitemShut
  {NoStop}%
\bibitem [{\citenamefont {Platz}\ \emph
  {et~al.}(2013{\natexlab{a}})\citenamefont {Platz}, \citenamefont
  {Forchheimer}, \citenamefont {Thol\'en},\ and\ \citenamefont
  {Haviland}}]{imafm4}%
  \BibitemOpen
  \bibfield  {author} {\bibinfo {author} {\bibfnamefont {D.}~\bibnamefont
  {Platz}}, \bibinfo {author} {\bibfnamefont {D.}~\bibnamefont {Forchheimer}},
  \bibinfo {author} {\bibfnamefont {E.~A.}\ \bibnamefont {Thol\'en}}, \ and\
  \bibinfo {author} {\bibfnamefont {D.~B.}\ \bibnamefont {Haviland}},\
  }\href@noop {} {\bibfield  {journal} {\bibinfo  {journal} {Nature Commun.}\
  }\textbf {\bibinfo {volume} {4}},\ \bibinfo {pages} {1360} (\bibinfo {year}
  {2013}{\natexlab{a}})}\BibitemShut {NoStop}%
\bibitem [{\citenamefont {Platz}\ \emph
  {et~al.}(2013{\natexlab{b}})\citenamefont {Platz}, \citenamefont
  {Forchheimer}, \citenamefont {Thol\'en},\ and\ \citenamefont
  {Haviland}}]{imafm5}%
  \BibitemOpen
  \bibfield  {author} {\bibinfo {author} {\bibfnamefont {D.}~\bibnamefont
  {Platz}}, \bibinfo {author} {\bibfnamefont {D.}~\bibnamefont {Forchheimer}},
  \bibinfo {author} {\bibfnamefont {E.~A.}\ \bibnamefont {Thol\'en}}, \ and\
  \bibinfo {author} {\bibfnamefont {D.~B.}\ \bibnamefont {Haviland}},\
  }\href@noop {} {\bibfield  {journal} {\bibinfo  {journal} {Beilstein J.
  Nanotechnol.}\ }\textbf {\bibinfo {volume} {4}},\ \bibinfo {pages} {45}
  (\bibinfo {year} {2013}{\natexlab{b}})}\BibitemShut {NoStop}%
\bibitem [{\citenamefont {Platz}\ \emph
  {et~al.}(2013{\natexlab{c}})\citenamefont {Platz}, \citenamefont
  {Forchheimer}, \citenamefont {Thol\'en},\ and\ \citenamefont
  {Haviland}}]{imafm6}%
  \BibitemOpen
  \bibfield  {author} {\bibinfo {author} {\bibfnamefont {D.}~\bibnamefont
  {Platz}}, \bibinfo {author} {\bibfnamefont {D.}~\bibnamefont {Forchheimer}},
  \bibinfo {author} {\bibfnamefont {E.~A.}\ \bibnamefont {Thol\'en}}, \ and\
  \bibinfo {author} {\bibfnamefont {D.~B.}\ \bibnamefont {Haviland}},\
  }\href@noop {} {\enquote {\bibinfo {title} {Polynomial force approximations
  and multifrequency atomic force microscopy},}\ }\bibinfo {howpublished}
  {e-print arXiv:cond-mat/1302.1829} (\bibinfo {year}
  {2013}{\natexlab{c}})\BibitemShut {NoStop}%
\bibitem [{\citenamefont {Raman}\ \emph {et~al.}(2008)\citenamefont {Raman},
  \citenamefont {Melcher},\ and\ \citenamefont {Tung}}]{gen1_cantDyn}%
  \BibitemOpen
  \bibfield  {author} {\bibinfo {author} {\bibfnamefont {A.}~\bibnamefont
  {Raman}}, \bibinfo {author} {\bibfnamefont {J.}~\bibnamefont {Melcher}}, \
  and\ \bibinfo {author} {\bibfnamefont {R.}~\bibnamefont {Tung}},\ }\href@noop
  {} {\bibfield  {journal} {\bibinfo  {journal} {Nano Today}\ }\textbf
  {\bibinfo {volume} {3}},\ \bibinfo {pages} {20} (\bibinfo {year}
  {2008})}\BibitemShut {NoStop}%
\bibitem [{\citenamefont {Rabe}\ \emph {et~al.}(1996)\citenamefont {Rabe},
  \citenamefont {Janser},\ and\ \citenamefont {Arnold}}]{gen2_cantVib}%
  \BibitemOpen
  \bibfield  {author} {\bibinfo {author} {\bibfnamefont {U.}~\bibnamefont
  {Rabe}}, \bibinfo {author} {\bibfnamefont {K.}~\bibnamefont {Janser}}, \ and\
  \bibinfo {author} {\bibfnamefont {W.}~\bibnamefont {Arnold}},\ }\href@noop {}
  {\bibfield  {journal} {\bibinfo  {journal} {Rev. Sci. Instrum.}\ }\textbf
  {\bibinfo {volume} {67}},\ \bibinfo {pages} {3281} (\bibinfo {year}
  {1996})}\BibitemShut {NoStop}%
\bibitem [{\citenamefont {Lee}\ \emph {et~al.}(2002)\citenamefont {Lee},
  \citenamefont {Howell}, \citenamefont {Raman},\ and\ \citenamefont
  {Reifenberger}}]{bern1}%
  \BibitemOpen
  \bibfield  {author} {\bibinfo {author} {\bibfnamefont {S.~I.}\ \bibnamefont
  {Lee}}, \bibinfo {author} {\bibfnamefont {S.~W.}\ \bibnamefont {Howell}},
  \bibinfo {author} {\bibfnamefont {A.}~\bibnamefont {Raman}}, \ and\ \bibinfo
  {author} {\bibfnamefont {R.}~\bibnamefont {Reifenberger}},\ }\href@noop {}
  {\bibfield  {journal} {\bibinfo  {journal} {Phys. Rev. B}\ }\textbf {\bibinfo
  {volume} {66}},\ \bibinfo {pages} {115409} (\bibinfo {year}
  {2002})}\BibitemShut {NoStop}%
\bibitem [{\citenamefont {Rodriguez}\ and\ \citenamefont
  {Garc\'ia}(2002)}]{bern2}%
  \BibitemOpen
  \bibfield  {author} {\bibinfo {author} {\bibfnamefont {T.~R.}\ \bibnamefont
  {Rodriguez}}\ and\ \bibinfo {author} {\bibfnamefont {R.}~\bibnamefont
  {Garc\'ia}},\ }\href@noop {} {\bibfield  {journal} {\bibinfo  {journal}
  {Appl. Phys. Lett.}\ }\textbf {\bibinfo {volume} {80}},\ \bibinfo {pages}
  {1646} (\bibinfo {year} {2002})}\BibitemShut {NoStop}%
\bibitem [{\citenamefont {Melcher}\ \emph {et~al.}(2007)\citenamefont
  {Melcher}, \citenamefont {Hu},\ and\ \citenamefont {Raman}}]{bern3}%
  \BibitemOpen
  \bibfield  {author} {\bibinfo {author} {\bibfnamefont {J.}~\bibnamefont
  {Melcher}}, \bibinfo {author} {\bibfnamefont {S.}~\bibnamefont {Hu}}, \ and\
  \bibinfo {author} {\bibfnamefont {A.}~\bibnamefont {Raman}},\ }\href@noop {}
  {\bibfield  {journal} {\bibinfo  {journal} {Appl. Phys. Lett.}\ }\textbf
  {\bibinfo {volume} {91}},\ \bibinfo {pages} {053101} (\bibinfo {year}
  {2007})}\BibitemShut {NoStop}%
\bibitem [{\citenamefont {Sader}(1998)}]{bern4}%
  \BibitemOpen
  \bibfield  {author} {\bibinfo {author} {\bibfnamefont {J.~E.}\ \bibnamefont
  {Sader}},\ }\href@noop {} {\bibfield  {journal} {\bibinfo  {journal} {J.
  Appl. Phys.}\ }\textbf {\bibinfo {volume} {84}},\ \bibinfo {pages} {64–}
  (\bibinfo {year} {1998})}\BibitemShut {NoStop}%
\bibitem [{\citenamefont {Love}(1888)}]{plate1_love}%
  \BibitemOpen
  \bibfield  {author} {\bibinfo {author} {\bibfnamefont {A.~E.~H.}\
  \bibnamefont {Love}},\ }\href@noop {} {\bibfield  {journal} {\bibinfo
  {journal} {Phil. Trans. R. Soc. A}\ }\textbf {\bibinfo {volume} {17}},\
  \bibinfo {pages} {491} (\bibinfo {year} {1888})}\BibitemShut {NoStop}%
\bibitem [{\citenamefont {Timoshenko}\ and\ \citenamefont
  {Woinowsky-Krieger}(1959)}]{plate2_timosh}%
  \BibitemOpen
  \bibfield  {author} {\bibinfo {author} {\bibfnamefont {S.}~\bibnamefont
  {Timoshenko}}\ and\ \bibinfo {author} {\bibfnamefont {S.}~\bibnamefont
  {Woinowsky-Krieger}},\ }\href@noop {} {\emph {\bibinfo {title} {Theory of
  plates and shells}}}\ (\bibinfo  {publisher} {McGraw–Hill New York},\
  \bibinfo {year} {1959})\BibitemShut {NoStop}%
\bibitem [{\citenamefont {Reddy}(2007)}]{plate3_review}%
  \BibitemOpen
  \bibfield  {author} {\bibinfo {author} {\bibfnamefont {J.~N.}\ \bibnamefont
  {Reddy}},\ }\href@noop {} {\emph {\bibinfo {title} {Theory and analysis of
  elastic plates and shells}}}\ (\bibinfo  {publisher} {CRC Press, Taylor and
  Francis},\ \bibinfo {year} {2007})\BibitemShut {NoStop}%
\bibitem [{\citenamefont {Reissner}\ and\ \citenamefont
  {Stein}(1951)}]{plate4_andtors}%
  \BibitemOpen
  \bibfield  {author} {\bibinfo {author} {\bibfnamefont {E.}~\bibnamefont
  {Reissner}}\ and\ \bibinfo {author} {\bibfnamefont {M.}~\bibnamefont
  {Stein}},\ }\href@noop {} {\emph {\bibinfo {title} {Torsion and transverse
  bending of cantilever plates}}},\ \bibinfo {type} {Tech. Rep.}\ \bibinfo
  {number} {2369}\ (\bibinfo  {institution} {National Advisory Committee for
  Aeronautics},\ \bibinfo {address} {Washington, D.C., USA},\ \bibinfo {year}
  {1951})\ \bibinfo {note} {technical Note}\BibitemShut {NoStop}%
\bibitem [{\citenamefont {Thol\'en}\ \emph {et~al.}(2011)\citenamefont
  {Thol\'en}, \citenamefont {Platz}, \citenamefont {Forchheimer}, \citenamefont
  {Thol\'en}, \citenamefont {Hutter},\ and\ \citenamefont
  {Haviland}}]{imafm_pr1}%
  \BibitemOpen
  \bibfield  {author} {\bibinfo {author} {\bibfnamefont {E.~A.}\ \bibnamefont
  {Thol\'en}}, \bibinfo {author} {\bibfnamefont {D.}~\bibnamefont {Platz}},
  \bibinfo {author} {\bibfnamefont {D.}~\bibnamefont {Forchheimer}}, \bibinfo
  {author} {\bibfnamefont {M.~O.}\ \bibnamefont {Thol\'en}}, \bibinfo {author}
  {\bibfnamefont {C.}~\bibnamefont {Hutter}}, \ and\ \bibinfo {author}
  {\bibfnamefont {D.~B.}\ \bibnamefont {Haviland}},\ }\href@noop {} {\bibfield
  {journal} {\bibinfo  {journal} {Rev. Sci. Instrum.}\ }\textbf {\bibinfo
  {volume} {82}},\ \bibinfo {pages} {026109} (\bibinfo {year}
  {2011})}\BibitemShut {NoStop}%
\bibitem [{ima()}]{imafm_pr2}%
  \BibitemOpen
  \href@noop {} {\enquote {\bibinfo {title} {Intermodulation products {AB}},}\
  }\bibinfo {howpublished}
  {\url{http://intermodulation-products.com}}\BibitemShut {NoStop}%
\bibitem [{\citenamefont {Derjaguin}\ \emph {et~al.}(1975)\citenamefont
  {Derjaguin}, \citenamefont {Muller},\ and\ \citenamefont
  {Toporov}}]{vdW-DMT}%
  \BibitemOpen
  \bibfield  {author} {\bibinfo {author} {\bibfnamefont {B.~V.}\ \bibnamefont
  {Derjaguin}}, \bibinfo {author} {\bibfnamefont {V.~M.}\ \bibnamefont
  {Muller}}, \ and\ \bibinfo {author} {\bibfnamefont {Y.~P.}\ \bibnamefont
  {Toporov}},\ }\href@noop {} {\bibfield  {journal} {\bibinfo  {journal} {J.
  Colloid Interface Sci.}\ }\textbf {\bibinfo {volume} {53}},\ \bibinfo {pages}
  {314} (\bibinfo {year} {1975})}\BibitemShut {NoStop}%
\bibitem [{\citenamefont {Garc\'ia}\ \emph
  {et~al.}(2006{\natexlab{b}})\citenamefont {Garc\'ia}, \citenamefont
  {G\'omez}, \citenamefont {Martinez}, \citenamefont {Patil}, \citenamefont
  {Dietz},\ and\ \citenamefont {Magerle}}]{vdW-DMT_mod1}%
  \BibitemOpen
  \bibfield  {author} {\bibinfo {author} {\bibfnamefont {R.}~\bibnamefont
  {Garc\'ia}}, \bibinfo {author} {\bibfnamefont {C.}~\bibnamefont {G\'omez}},
  \bibinfo {author} {\bibfnamefont {N.}~\bibnamefont {Martinez}}, \bibinfo
  {author} {\bibfnamefont {S.}~\bibnamefont {Patil}}, \bibinfo {author}
  {\bibfnamefont {C.}~\bibnamefont {Dietz}}, \ and\ \bibinfo {author}
  {\bibfnamefont {R.}~\bibnamefont {Magerle}},\ }\href@noop {} {\bibfield
  {journal} {\bibinfo  {journal} {Phys. Rev. Lett.}\ }\textbf {\bibinfo
  {volume} {97}},\ \bibinfo {pages} {016103} (\bibinfo {year}
  {2006}{\natexlab{b}})}\BibitemShut {NoStop}%
\bibitem [{\citenamefont {Gotsmann}\ \emph {et~al.}(1999)\citenamefont
  {Gotsmann}, \citenamefont {Seidel}, \citenamefont {Anczykowski},\ and\
  \citenamefont {Fuchs}}]{vdW-DMT_mod2}%
  \BibitemOpen
  \bibfield  {author} {\bibinfo {author} {\bibfnamefont {B.}~\bibnamefont
  {Gotsmann}}, \bibinfo {author} {\bibfnamefont {C.}~\bibnamefont {Seidel}},
  \bibinfo {author} {\bibfnamefont {B.}~\bibnamefont {Anczykowski}}, \ and\
  \bibinfo {author} {\bibfnamefont {H.}~\bibnamefont {Fuchs}},\ }\href@noop {}
  {\bibfield  {journal} {\bibinfo  {journal} {Phys. Rev. B}\ }\textbf {\bibinfo
  {volume} {60}},\ \bibinfo {pages} {11051} (\bibinfo {year}
  {1999})}\BibitemShut {NoStop}%
\bibitem [{\citenamefont {Melcher}\ \emph {et~al.}(2008)\citenamefont
  {Melcher}, \citenamefont {Hu},\ and\ \citenamefont {Raman}}]{vdW-DMT_mod3}%
  \BibitemOpen
  \bibfield  {author} {\bibinfo {author} {\bibfnamefont {J.}~\bibnamefont
  {Melcher}}, \bibinfo {author} {\bibfnamefont {S.}~\bibnamefont {Hu}}, \ and\
  \bibinfo {author} {\bibfnamefont {A.}~\bibnamefont {Raman}},\ }\href@noop {}
  {\bibfield  {journal} {\bibinfo  {journal} {Rev. Sci. Instrum.}\ }\textbf
  {\bibinfo {volume} {79}},\ \bibinfo {pages} {061301} (\bibinfo {year}
  {2008})}\BibitemShut {NoStop}%
\bibitem [{Note1()}]{Note1}%
  \BibitemOpen
  \bibinfo {note} {Contrary to the model (\ref {poly_general}), one might
  consider a model which at first glance appears more suitable, where the force
  is in the form of a product of single variable polynomials $\protect
  \mathaccentV {tilde}07EF_{i}(q_1,\protect \dots ,q_N) = \DOTSB \prod@
  \slimits@ _{m=1}^N P_m(q_m)$. While this model has a much smaller total
  number of parameters to determine, upon insertion into (\ref {errorf}), we
  encounter two principal difficulties: (i) if we explicitly perform
  multiplication and then take the Fourier transform, we obtain a system for
  the unknown parameters which is nonlinear in the parameters; (ii) if we
  insert it as it is, the deconvolution problem must be solved $\protect
  \mathaccentV {hat}05EF_i=\protect \mathaccentV {hat}05EP_1\ast \protect \dots
  \ast \protect \mathaccentV {hat}05EP_N$ which requires knowledge of the
  spectral components outside narrow bands surrounding resonances.}\BibitemShut
  {Stop}%
\bibitem [{Note2()}]{Note2}%
  \BibitemOpen
  \bibinfo {note} {Strictly speaking, the size of the system is $2B_{i}$ as the
  Fourier transform of a real function is symmetrical with respect to the zero
  frequency but complex conjugated. However, this fact does not provide any
  additional information and can be used only for improving numerical stability
  of calculations. Solving this system separately for real and imaginary parts
  gives the same value of $\protect \mathbf {g}$.}\BibitemShut {Stop}%
\bibitem [{\citenamefont {Jetter}\ \emph {et~al.}(2006)\citenamefont {Jetter},
  \citenamefont {Buhmann}, \citenamefont {Haussmann}, \citenamefont
  {Schaback},\ and\ \citenamefont {Stoeckler}}]{interpol1}%
  \BibitemOpen
  \bibinfo {editor} {\bibfnamefont {K.}~\bibnamefont {Jetter}}, \bibinfo
  {editor} {\bibfnamefont {M.}~\bibnamefont {Buhmann}}, \bibinfo {editor}
  {\bibfnamefont {W.}~\bibnamefont {Haussmann}}, \bibinfo {editor}
  {\bibfnamefont {R.}~\bibnamefont {Schaback}}, \ and\ \bibinfo {editor}
  {\bibfnamefont {J.}~\bibnamefont {Stoeckler}},\ eds.,\ \href@noop {} {\emph
  {\bibinfo {title} {Topics in Multivariate Approximation and
  Interpolation}}},\ \bibinfo {series} {Studies in Computational Mathematics},
  Vol.~\bibinfo {volume} {12}\ (\bibinfo  {publisher} {Elsevier Science},\
  \bibinfo {year} {2006})\BibitemShut {NoStop}%
\bibitem [{\citenamefont {Schlömilch}\ \emph {et~al.}(1901)\citenamefont
  {Schlömilch}, \citenamefont {Witzschel}, \citenamefont {Cantor},
  \citenamefont {Kahl}, \citenamefont {Mehmke},\ and\ \citenamefont
  {Runge}}]{interpol2}%
  \BibitemOpen
  \bibfield  {author} {\bibinfo {author} {\bibfnamefont {O.~X.}\ \bibnamefont
  {Schlömilch}}, \bibinfo {author} {\bibfnamefont {B.}~\bibnamefont
  {Witzschel}}, \bibinfo {author} {\bibfnamefont {M.}~\bibnamefont {Cantor}},
  \bibinfo {author} {\bibfnamefont {E.}~\bibnamefont {Kahl}}, \bibinfo {author}
  {\bibfnamefont {R.}~\bibnamefont {Mehmke}}, \ and\ \bibinfo {author}
  {\bibfnamefont {C.}~\bibnamefont {Runge}},\ }\href@noop {} {\bibfield
  {journal} {\bibinfo  {journal} {Z. Math. Phys.}\ }\textbf {\bibinfo {volume}
  {46}},\ \bibinfo {pages} {224–} (\bibinfo {year} {1901})}\BibitemShut
  {NoStop}%
\bibitem [{\citenamefont {Dahlquist}\ and\ \citenamefont
  {Bj{\"o}rk}(1974)}]{interpol3}%
  \BibitemOpen
  \bibfield  {author} {\bibinfo {author} {\bibfnamefont {G.}~\bibnamefont
  {Dahlquist}}\ and\ \bibinfo {author} {\bibfnamefont {A.}~\bibnamefont
  {Bj{\"o}rk}},\ }in\ \href@noop {} {\emph {\bibinfo {booktitle} {Numerical
  Methods}}},\ \bibinfo {series and number} {Dover Books on Mathematics}\
  (\bibinfo  {publisher} {Dover Publications Inc.},\ \bibinfo {year} {1974})\
  \bibinfo {type} {Section}\ \bibinfo {chapter} {4.3.4}, pp.\ \bibinfo {pages}
  {101--–103}\BibitemShut {NoStop}%
\bibitem [{Note3()}]{Note3}%
  \BibitemOpen
  \bibinfo {note} {In principle, it can be generalized for case of $N$
  collinear modes $\chi = \DOTSB \sum@ \slimits@ _{i=0}^{N}\chi
  _i$.}\BibitemShut {Stop}%
\bibitem [{\citenamefont {Sahin}\ \emph {et~al.}(2004)\citenamefont {Sahin},
  \citenamefont {Yaralioglu}, \citenamefont {Grow}, \citenamefont {Zappe},
  \citenamefont {Atalar}, \citenamefont {Quate},\ and\ \citenamefont
  {Solgaard}}]{harm_cant1}%
  \BibitemOpen
  \bibfield  {author} {\bibinfo {author} {\bibfnamefont {O.}~\bibnamefont
  {Sahin}}, \bibinfo {author} {\bibfnamefont {G.}~\bibnamefont {Yaralioglu}},
  \bibinfo {author} {\bibfnamefont {R.}~\bibnamefont {Grow}}, \bibinfo {author}
  {\bibfnamefont {S.~F.}\ \bibnamefont {Zappe}}, \bibinfo {author}
  {\bibfnamefont {A.}~\bibnamefont {Atalar}}, \bibinfo {author} {\bibfnamefont
  {C.}~\bibnamefont {Quate}}, \ and\ \bibinfo {author} {\bibfnamefont
  {O.}~\bibnamefont {Solgaard}},\ }\href@noop {} {\bibfield  {journal}
  {\bibinfo  {journal} {Sens. Actuators A}\ }\textbf {\bibinfo {volume}
  {114}},\ \bibinfo {pages} {183} (\bibinfo {year} {2004})}\BibitemShut
  {NoStop}%
\bibitem [{\citenamefont {Hindmarsh}\ \emph {et~al.}(2005)\citenamefont
  {Hindmarsh}, \citenamefont {Brown}, \citenamefont {Grant}, \citenamefont
  {Lee}, \citenamefont {Serban}, \citenamefont {Shumaker},\ and\ \citenamefont
  {Woodward}}]{CVODE}%
  \BibitemOpen
  \bibfield  {author} {\bibinfo {author} {\bibfnamefont {A.~C.}\ \bibnamefont
  {Hindmarsh}}, \bibinfo {author} {\bibfnamefont {P.~N.}\ \bibnamefont
  {Brown}}, \bibinfo {author} {\bibfnamefont {K.~E.}\ \bibnamefont {Grant}},
  \bibinfo {author} {\bibfnamefont {S.~L.}\ \bibnamefont {Lee}}, \bibinfo
  {author} {\bibfnamefont {R.}~\bibnamefont {Serban}}, \bibinfo {author}
  {\bibfnamefont {D.~E.}\ \bibnamefont {Shumaker}}, \ and\ \bibinfo {author}
  {\bibfnamefont {C.~S.}\ \bibnamefont {Woodward}},\ }\href@noop {} {\bibfield
  {journal} {\bibinfo  {journal} {Trans. Math. Software}\ }\textbf {\bibinfo
  {volume} {31}},\ \bibinfo {pages} {363} (\bibinfo {year} {2005})}\BibitemShut
  {NoStop}%
\bibitem [{\citenamefont {Lozano}\ \emph {et~al.}(2010)\citenamefont {Lozano},
  \citenamefont {Kiracofe}, \citenamefont {Melcher}, \citenamefont {Garc\'ia},\
  and\ \citenamefont {Raman}}]{calib1}%
  \BibitemOpen
  \bibfield  {author} {\bibinfo {author} {\bibfnamefont {J.~R.}\ \bibnamefont
  {Lozano}}, \bibinfo {author} {\bibfnamefont {D.}~\bibnamefont {Kiracofe}},
  \bibinfo {author} {\bibfnamefont {J.}~\bibnamefont {Melcher}}, \bibinfo
  {author} {\bibfnamefont {R.}~\bibnamefont {Garc\'ia}}, \ and\ \bibinfo
  {author} {\bibfnamefont {A.}~\bibnamefont {Raman}},\ }\href@noop {}
  {\bibfield  {journal} {\bibinfo  {journal} {Nanotechnology}\ }\textbf
  {\bibinfo {volume} {21}},\ \bibinfo {pages} {465502} (\bibinfo {year}
  {2010})}\BibitemShut {NoStop}%
\bibitem [{\citenamefont {Green}\ \emph {et~al.}(2004)\citenamefont {Green},
  \citenamefont {Lioe}, \citenamefont {Cleveland}, \citenamefont {Proksch},
  \citenamefont {Mulvaney},\ and\ \citenamefont {Sader}}]{calib2}%
  \BibitemOpen
  \bibfield  {author} {\bibinfo {author} {\bibfnamefont {C.~P.}\ \bibnamefont
  {Green}}, \bibinfo {author} {\bibfnamefont {H.}~\bibnamefont {Lioe}},
  \bibinfo {author} {\bibfnamefont {J.~P.}\ \bibnamefont {Cleveland}}, \bibinfo
  {author} {\bibfnamefont {R.}~\bibnamefont {Proksch}}, \bibinfo {author}
  {\bibfnamefont {P.}~\bibnamefont {Mulvaney}}, \ and\ \bibinfo {author}
  {\bibfnamefont {J.~E.}\ \bibnamefont {Sader}},\ }\href@noop {} {\bibfield
  {journal} {\bibinfo  {journal} {Rev. Sci. Instrum.}\ }\textbf {\bibinfo
  {volume} {75}},\ \bibinfo {pages} {1988} (\bibinfo {year}
  {2004})}\BibitemShut {NoStop}%
\bibitem [{\citenamefont {Feiler}\ \emph {et~al.}(2000)\citenamefont {Feiler},
  \citenamefont {Attard},\ and\ \citenamefont {Larson}}]{calib3}%
  \BibitemOpen
  \bibfield  {author} {\bibinfo {author} {\bibfnamefont {A.}~\bibnamefont
  {Feiler}}, \bibinfo {author} {\bibfnamefont {P.}~\bibnamefont {Attard}}, \
  and\ \bibinfo {author} {\bibfnamefont {I.}~\bibnamefont {Larson}},\
  }\href@noop {} {\bibfield  {journal} {\bibinfo  {journal} {Rev. Sci.
  Instrum.}\ }\textbf {\bibinfo {volume} {71}},\ \bibinfo {pages} {2746}
  (\bibinfo {year} {2000})}\BibitemShut {NoStop}%
\end{thebibliography}%

\end{document}